\documentclass[aps,pre,floatfix,twocolumn,10pt,superscriptaddress]{revtex4-1}
\pdfoutput=1
\usepackage{natbib}
\usepackage{amssymb}
\usepackage{amsfonts}
\usepackage{amsmath}
\usepackage[english]{babel}
\usepackage[T1]{fontenc}
\usepackage{lmodern}
\usepackage{graphicx}
\usepackage{hyperref}
\hypersetup{%
colorlinks=true,%
citecolor=blue,%
filecolor=blue,%
linkcolor=blue,%
urlcolor=blue,%
pdfauthor={V. H. Purrello, V. Lecomte, J. L. Iguain, A. B. Kolton},%
pdftitle={Hysteretic depinning of a particle in a periodic potential}
}

\newcommand{\dd}{\text{d}}
\newcommand{\deltaf}{\delta\!f}
\newcommand{\deltam}{\delta m}

\begin{document}

\author{V\'ictor H. Purrello}
\email{vpurrello@ifimar-conicet.gob.ar}
\author{Jos\'e L. Iguain}
\email{iguain@mdp.edu.ar}
\affiliation{IFIMAR, Facultad de Ciencias Exactas y Naturales,
Universidad Nacional de Mar del Plata, CONICET, 7600 Mar del Plata,
Argentina}

\author{Vivien Lecomte}
\email{vivien.lecomte@univ-grenoble-alpes.fr}
\affiliation{ Universit\'e Grenoble Alpes, CNRS, LIPhy, 38000 Grenoble, France}

\author{Alejandro B. Kolton}
\email{alejandro.kolton@cab.cnea.gov.ar}
\affiliation{Centro At\'omico Bariloche and Instituto Balseiro,
CNEA, CONICET and Universidad Nacional de Cuyo, 8400 Bariloche, Argentina}

\title{Hysteretic depinning of a particle in a periodic potential: \\
Phase diagram and criticality}

\begin{abstract}
We consider a massive particle driven with a constant force in a periodic potential and subjected to
a dissipative friction.
As a function of the drive and damping, the phase diagram of this paradigmatic model is well known
to present a pinned, a sliding, and a bistable regime separated by three distinct bifurcation lines.
In physical terms, the average velocity $v$ of the particle is nonzero only if either
(i) the driving force is large enough to remove any stable point,
forcing the particle to slide,
or
(ii) there are local minima but the damping is small enough,
below a critical damping, for the inertia to allow the particle to cross barriers and follow a limit
cycle; this regime is bistable and whether $v>0$ or $v=0$ depends on the initial state.
In this paper, we focus on the asymptotes of the critical line separating the bistable and the
pinned regimes.
First, we study its behavior near the ``triple point'' where the pinned, the bistable, and the
sliding dynamical regimes meet.
Just below the critical damping we uncover a critical regime, where the line approaches the triple
point following a power-law behavior.
We show that its exponent is controlled by the normal form of the tilted potential close to its
critical force.
Second, in the opposite regime of very low damping, we revisit existing results by providing a
simple method to determine analytically the exact behavior of the line in the case of a generic
potential.
The analytical estimates, accurately confirmed numerically, are obtained by exploiting exact soliton
solutions describing the orbit in a modified tilted potential which can be mapped to the original
tilted washboard potential.
Our methods and results are  particularly useful for an accurate description of underdamped
nonuniform oscillators driven near their triple point.
\end{abstract}

\maketitle

\section{Introduction}
\label{sec:inercial_d0}
Let $x(t)$ be the one-dimensional position of an underdamped particle driven in a generic
differentiable periodic potential $V(x)$ with spatial period $\ell$, described by the deterministic
equation of motion
\begin{equation}
  m \ddot x + \gamma \dot x = -V'(x) + f \:,
\label{eq:mainequation}
\end{equation}
where $m$ is the mass of the particle, $\gamma$ a friction constant corresponding to a kinetic
friction force proportional to the instantaneous velocity, and $f>0$ is a constant driving force.
Equation~\eqref{eq:mainequation} is a ubiquitous differential equation.
It provides both a textbook example of  bifurcations in two-dimensional nonlinear systems
(e.g., see Ref.~\cite{strogatz2018nonlinear}) and a useful model for large number of concrete
physical
systems, such as nonuniform oscillators.
Already the simple case $V(x) \propto \cos(2\pi x/\ell)$ describes both the simple pendulum driven
by a constant torque and the underdamped Josephson junction driven by a constant external electric
current.
In the latter example, $x(t)$ represents the superconducting order parameter phase difference
across a small junction separating two superconducting regions with a capacitance $C \propto m$ and
electric resistance~$R \propto \gamma^{-1}$, with all these ideal elements effectively connected in
parallel in the so-called Stewart--McCumber model~\cite{tinkham2004introduction}.
Note that the steady-state time-averaged velocity $v\equiv \langle \dot{x} \rangle$ as a function
of $f$ models the voltage-current characteristics of the junction.
Many of the properties predicted from Eq.~\eqref{eq:mainequation} have been observed experimentally
in these superconducting devices~\cite{baronepaterno}.

The phase diagram of the large-time behavior of solutions to Eq.~\eqref{eq:mainequation} can be
solved analytically for $m=0$, i.e., in the overdamped case.
We review here its derivation for comparison to the inertial case.
The steady state is entirely determined by whether the tilted potential $V(x)-f x$ presents
barriers.
A continuous depinning transition exists at the unique threshold force
$f_c^0 = \max_xV'(x) = V'(x^*)$ at which barriers vanish when increasing the drive $f$ from $0$.
Indeed, below $f_c^0$, barriers exist and the damping pins the particle in a local minimum at large
time; the average velocity is 0.
At $f_c^0$, a saddle-node bifurcation occurs.
Above $f_c^0$, the instantaneous velocity becomes periodic in time:
$\dot{x}(t)=\dot{x}(t+\tau)$, with a  positive average value $v$.
We have that  $v \sim (f-f_c^0)^\beta$ if $0 < f - f_c^0 \ll f_c^0$,
and $v \approx f/\gamma$ if $f \gg f_c^0$.
The so-called depinning exponent $\beta \geq 0$ depends on the normal form of the saddle-node
bifurcation at $f_c^0$.
For typical analytical potentials such that $f-V'(x) \approx (f-f_c^0)+k|x-x^*|^2$
for $|f-f_c^0| \ll f_c^0$ with a constant $k>0$ and $|x-x^*| \ll \ell$, we have the well known
square-root depinning law with $\beta=1/2$.
It is derived as follows: On a time period $\tau$, the trajectory along the limit cycle spends most
of its time close to the bottleneck at $x = x^*$, and thus
$\tau = \int_0^\ell dx/[f-V'(x)] \approx \int_0^\ell dx/[(f-f_c^0)+k|x-x^*|^2] \sim
(f-f_c^0)^{-1/2}$.
Therefore, since the particle travels a distance $\ell$ during the time $\tau$,
one has $v = \ell / \tau \sim (f-f_c^0)^{1/2}$ just above $f_c^0$.
The exponent  $\beta$ depends on the behavior of $V(x)$ in the vicinity of the bottleneck.
More generically,
the normal form of the saddle-node bifurcation is controlled by an expansion of the form
$f-V'(x) \approx (f-f_c^0)+k|x-x^*|^\Upsilon$,
and we obtain $\beta=1-1/\Upsilon$, for $\Upsilon>1$~\cite{purrello2017}.
Furthermore, the $m=0$ case of Eq.~\eqref{eq:mainequation} can be solved analytically even in the
presence of an additive thermal Langevin noise.
Analytical expressions for the thermally averaged velocity $v$ for general $V(x)$ can be obtained
solving the Fokker--Planck equation for the steady-state
probability~\cite{stratonovich_oscillator_1965,ambegaokar1969,ledoussal_1995,scheidl_mobility_1995,risken_fokker-planck_1996}.
At finite temperature, the zero-temperature velocity-force characteristics is rounded around
$f_c^0$ (e.g., see Ref.~\cite{purrello2017}), and $v$ is positive and finite for any $f>0$, as
thermal activation can help to overcome barriers when $0 \leq f<f_c^0$.
Nevertheless, at small temperatures the fingerprint of the $T=0$ and $f=f_c^0$ depinning transition
is present and an analogy with continuous equilibrium phase transitions can be drawn, with $v$
representing the order parameter~\cite{bishop1978}.

\begin{figure}[tb]
\centering
\includegraphics[width=.9\columnwidth]{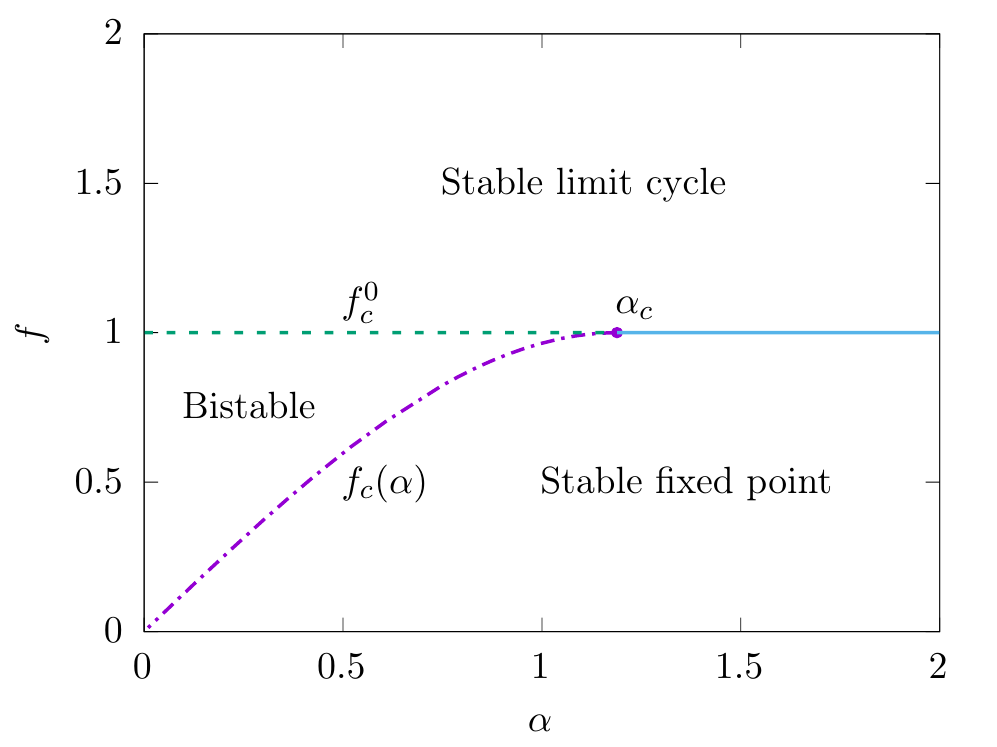}
\caption{
Phase diagram of Eq.~\eqref{eq:mainequation} for a driven particle in the washboard
potential $V(x) = -\cos(x)$, in the large-time asymptotics.
Control parameters are the driving force $f>0$ and the damping parameter
$\alpha\equiv\gamma/\sqrt{m}$.
In the stable limit cycle domain, the mean velocity $v$ of the particle is finite
(and the motion is periodic), while it is zero in the stable fixed point domain
(the particle is pinned in a local minimum).
In the bistable regime, whether $v=0$ of $v>0$ depends on the initial conditions.
The magenta dash-dotted line $f_c(\alpha)$ represents a homoclinic bifurcation,
the green dashed line a finite-period saddle-node bifurcation, while
the solid light blue line is an infinite period bifurcation.
The lines are obtained using standard numerical integration methods for Eq.~\eqref{eq:mainequation}.
In this paper we obtain analytical expressions for $f_c(\alpha)$ in the $\alpha \to 0$ limit for an
arbitrary pinning potential $V(x)$, and we describe the universal power-law behavior of
$f_c(\alpha)$ in the  $\alpha \to \alpha_c$ limit
when approaching the triple point where the three bifurcation lines meet.
}
\label{fig:phase_diag}
\end{figure}

\begin{figure}[tb]
\centering
\includegraphics[width=.9\columnwidth]{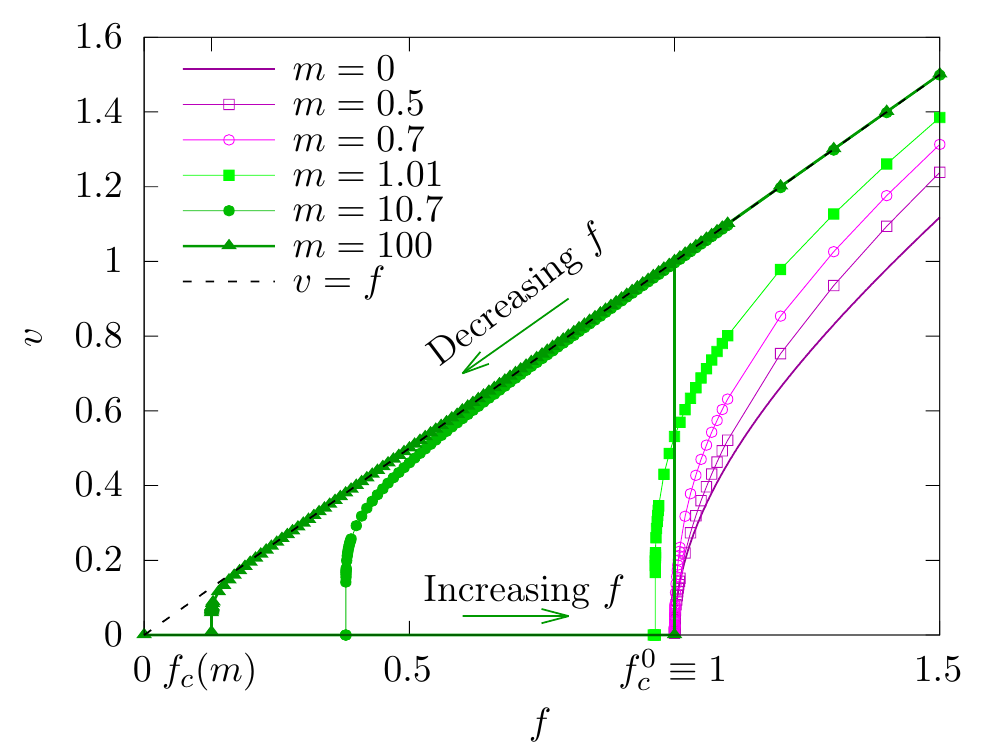}
\caption{
Hysteresis loops in the velocity-force characteristics for $V(x)=-\cos(x)$ in
Eq.~\eqref{eq:mainequation}.
For high damping, corresponding to masses $m<m_c$ at fixed $\gamma$, with $\gamma=1$ in this case,
the particle depins at $f_c^0\equiv1$ (magenta lines with empty points), without displaying
hysteresis.
For masses $m>m_c$ (green lines with filled points), there is a hysteretic depinning:
Slowly increasing the applied force from $f=0$, the particle depins at $f_c^0$,
acquiring a finite steady-state velocity only above it.
Then, slowly decreasing $f$ from above $f_c^0$, the particle gets pinned only at the critical force
$f_c(m)<f_c^0$.
In this paper we analytically calculate the hysteretic range $[f_c^0-f_c(m)]$ for masses close to
$m_c$.
}
\label{fig:v_vs_f}
\end{figure}

Compared to the overdamped case, far fewer analytical results are known for the inertial dynamics
[$m>0$ in Eq.~\eqref{eq:mainequation}].
Now the phase diagram depends not only on the properties of the tilted potential $V(x)-fx$ but also
on the relative values of the mass $m$ and the friction $\gamma$.
For the paradigmatic case $V(x)=-\cos(x)$, the representative phase diagram as a function of the
damping parameters $\alpha = \gamma/\sqrt{m}$ and $f$ is qualitatively well
known~\cite{levi1978dynamics,strogatz2018nonlinear}, and shown in Fig.~\ref{fig:phase_diag}.
It is characterized by three bifurcations lines meeting at a triple point.
As in the zero-mass case, the line $f=f_c^0$ delimitates the region where the tilted potential has
local minima ($f<f_c^0$) or not ($f\geq f_c^0$).
For $\alpha>\alpha_c$, the damping is strong and having a zero velocity or not
depends only on this intrinsic property of the tilted potential:
The depinning transition occurs at $f_c^0$.
It is described by an infinite-period bifurcation where the stable fixed point annihilates with the
unstable fixed point and disappears as $f\uparrow f_c^0$.
For $f>f_c^0$, the existence of a stable limit cycle implies $v>0$, and as in the overdamped case,
$v \sim (f-f_c^0)^{1/2}$ (though with a nontrivial $\alpha$-dependent prefactor).
As $\alpha\to\infty$, one recovers the zero-mass case.

In contrast, for $\alpha<\alpha_c$, the damping is low enough to allow a bistable regime when
$f_c(\alpha)<f<f_c^0$.
In that regime, there is a coexistence between a stable point (a minimum of the tilted potential)
for which $v=0$,
and limit cycle (where inertia allows the particle to cross barriers) for which $v>0$.
Initial conditions determine whether $v=0$ or $v>0$, and varying $f$ slowly leads to hysteresis
(see Fig.~\ref{fig:v_vs_f}).
The phase-space trajectories of Fig.~\ref{fig:phasespaceorbits} illustrate the possible orbits in
each regime.
Note that the transition line $f_c(\alpha)$ for $\alpha<\alpha_c$ is described by a homoclinic
bifurcation, for which the depinning transition occurs as $v\sim 1/|\log [f-f_c(\alpha)]|$ as
$f\downarrow f_c(\alpha)$.
Such logarithmic behavior, corresponding to a ``0'' depinning exponent $\beta$, is in sharp contrast
with the $\beta>0$ exponent that governs the transition for $\alpha>\alpha_c$.
Although the phase diagram of Fig.~\ref{fig:phase_diag} is obtained for the particular case
$V(x)=-\cos(x)$, it is qualitatively the same for any smooth periodic potential presenting a single minimum and maximum, as we show in this
paper.

\section{Summary of the results}
\label{sec:summary-results}
We focus on the behavior of the homoclinic bifurcation line, which separates the stable fixed point
regime from the bistable one (see Fig.~\ref{fig:v_vs_f}).
We will use two complementary view points: either varying directly the damping constant
$\alpha=\gamma/\sqrt{m}$, or,
at fixed friction $\gamma$, varying the mass $m$.
The critical line will respectively be denoted by $f_c(\alpha)$ or $f_c(m)$.
At fixed $\gamma$, to the critical damping $\alpha_c$ corresponds a
critical mass $m_c=(\gamma/\alpha_c)^2$, above which inertial effects matter.

As seen on Fig.~\ref{fig:phase_diag},
on the one hand $f_c(\alpha)$ tends to zero when $\alpha \to 0$ (i.e., in the weakly damped limit)
while it goes smoothly to $f_c^0$, as $\alpha \to \alpha_c$.
For $V(x)=-\cos(x)$, Guckenheimer and Holmes used Melnikov's technique~\cite{melnikov1963}
to show that $f_c(\alpha) \sim 4\alpha/\pi$ as $\alpha \to 0$~\cite{guckenheimer1983}.
Rather surprisingly, to the best of our knowledge, no analytical prediction regarding the way
$f_c(\alpha)$ approaches $f_c^0$ as $\alpha \to \alpha_c$ nor any expression for $\alpha_c$ were
reported.

Yet, as we show, the line $f_c(\alpha)$ presents universal and scaling properties which are
important to understand if we wish to use Eq.~\eqref{eq:mainequation} as a toy model for more
complex extended systems with inertia.
These were not reported neither and in this work we provide elements to fill this gap.
To do so, we employ a method to derive analytically $f_c^0-f_c(\alpha)$ as a function of
$\alpha$ close to the triple point ($\alpha-\alpha_c \ll \alpha_c$),
including explicit expressions for $\alpha_c$.
We generalize the $\alpha \to 0$ asymptotics  $f_c(\alpha) \sim 4\alpha/\pi$ obtained by
Guckenheimer and Holmes~\cite{guckenheimer1983} for the cosine potential to the case of an arbitrary
potential.
We also show that while the $\alpha \to 0$ asymptotic scaling $f_c(\alpha) \propto \alpha$ is rather
insensible to the details of the periodic potential, the scaling behavior when $\alpha \to \alpha_c$
is of the form $f_c^0-f_c(\alpha) \sim (\alpha_c-\alpha)^\delta$,
where the exponent  $\delta>0$ depends on the normal form of the bifurcation at $f_c^0$ for
$\alpha>\alpha_c$.
If, near the saddle-node bifurcation point $x^*$, the force can be expanded as
$f-V'(x) \approx (f-f_c^0)+k|x-x^*|^\Upsilon+\ldots$, with $k$ a positive constant and $\Upsilon>1$,
we obtain that $\delta=\Upsilon$.
In particular, for the cosine potential $V(x)=-\cos(x)$, which corresponds to the paradigmatic
pendulum and Josephson-junction problems, we obtain that the homoclinic line is characterized by the
scaling $f_c^0-f_c(\alpha) \sim (\alpha_c-\alpha)^\delta$, with $\delta=2$.

We also show that the values of $\alpha_c$, $f_c(\alpha)$ and of the prefactor of the scaling laws
are nonuniversal but depend on the details of $V(x)$ that are relevant for a precise interplay
among dissipation, inertia, and drive.
They can nevertheless be also estimated analytically.
To obtain these results we  exploit the fact that, associated to the homoclinic bifurcation at
$f_c(\alpha)$, there exists a ``critical trajectory,'' or homoclinic orbit, that connects a local
maximum of the tilted potential to the next one, on an infinite time window and with $\dot x\to 0$
at both extremal points [see Fig.~\ref{fig:phasespaceorbits}\,(b)].
This homoclinic orbit cannot be obtained analytically in general, and presents no obvious scaling
form.
To study it in spite of these issues, we use a tilted periodic potential $\mathcal{V}_F(x)$,
different from the original one $V(x)-f x$, but which has the advantage that the homoclinic orbit
at $f_c(\alpha)$ can be found exactly (it is a dissipative soliton).
We then map the exact critical properties derived for the effective potential to the ones of the
original tilted potential $V(x)-f x$, to estimate $f_c(\alpha)$ in the regimes of interest.
We show that the procedure is quite general and applies to various cases.

For the standard $\Upsilon=2$ case, we find
\begin{align}
f_c(\alpha) &\sim \mathcal N \alpha,
  & [\alpha \to 0]
\\
f_c^0 - f_c(\alpha) &\sim(\alpha_c - \alpha)^2,
  & [\alpha_c-\alpha \to 0^+]
\\
v &\sim (f-f_c^0)^{1/2},
  & [\alpha>\alpha_c, f - f_c^0 \to 0^+]
\end{align}
with an exact determination of the prefactor $\mathcal N$ [see Eq.~\eqref{eq:relationmcdefN}].
In the more general case of an arbitrary value of the exponent $\Upsilon$, one finds
\begin{align}
f_c(\alpha) &\sim \mathcal N \alpha,
  & \alpha \to 0
\\
f_c^0 - f_c(\alpha) &\sim (\alpha_c - \alpha)^\Upsilon,
  & [\alpha_c-\alpha \to 0^+]
\\
v &\sim (f-f_c^0)^{1-\frac{1}{\Upsilon}},
  & [\alpha>\alpha_c, f - f_c^0 \to 0^+]
\end{align}
with the same expression for the prefactor $\mathcal N$.

\section{Organization of the paper}
\label{sec:orgpaper}
In Sec.~\ref{sec:crittraj}, we review the properties of the critical trajectory which separates
static from running solutions in the bistable regime.
Then in Sec.~\ref{sec:def-pot-phi4}, we present a particular periodic potential,
along with also a particular way of tilting it,
from which $f_c(\alpha)$ for $\alpha<\alpha_c$, and $\alpha_c$ can be determined analytically.
We show how to use the results obtained from the modified tilted potential in order to estimate
these properties for the paradigmatic case $V(x) \propto -\cos(x)$.
Scaling forms and characteristic quantities, are discussed.
In Sec.~\ref{sec:genupsilon}, we generalize the approach for the more general case
$1 < \Upsilon \neq 2$, and discuss the universality of the different results.
In Section~\ref{sec:gener-scal-theory}, we review and generalize the large-damping approach of
Guckenheimer and Holmes, which allows us to determine exactly $f_c(\alpha)$ in the regime
$\alpha\ll\alpha_c$ for a generic potential.
Section~\ref{sec:0d_numerics} presents numerical validation of our predictions together with
additional observations.
Section~\ref{sec:conclusions} contains our conclusions and perspectives.

\section{The critical trajectory}
\label{sec:crittraj}
\begin{figure}[tb]
\centering
\includegraphics[width=.9\columnwidth]{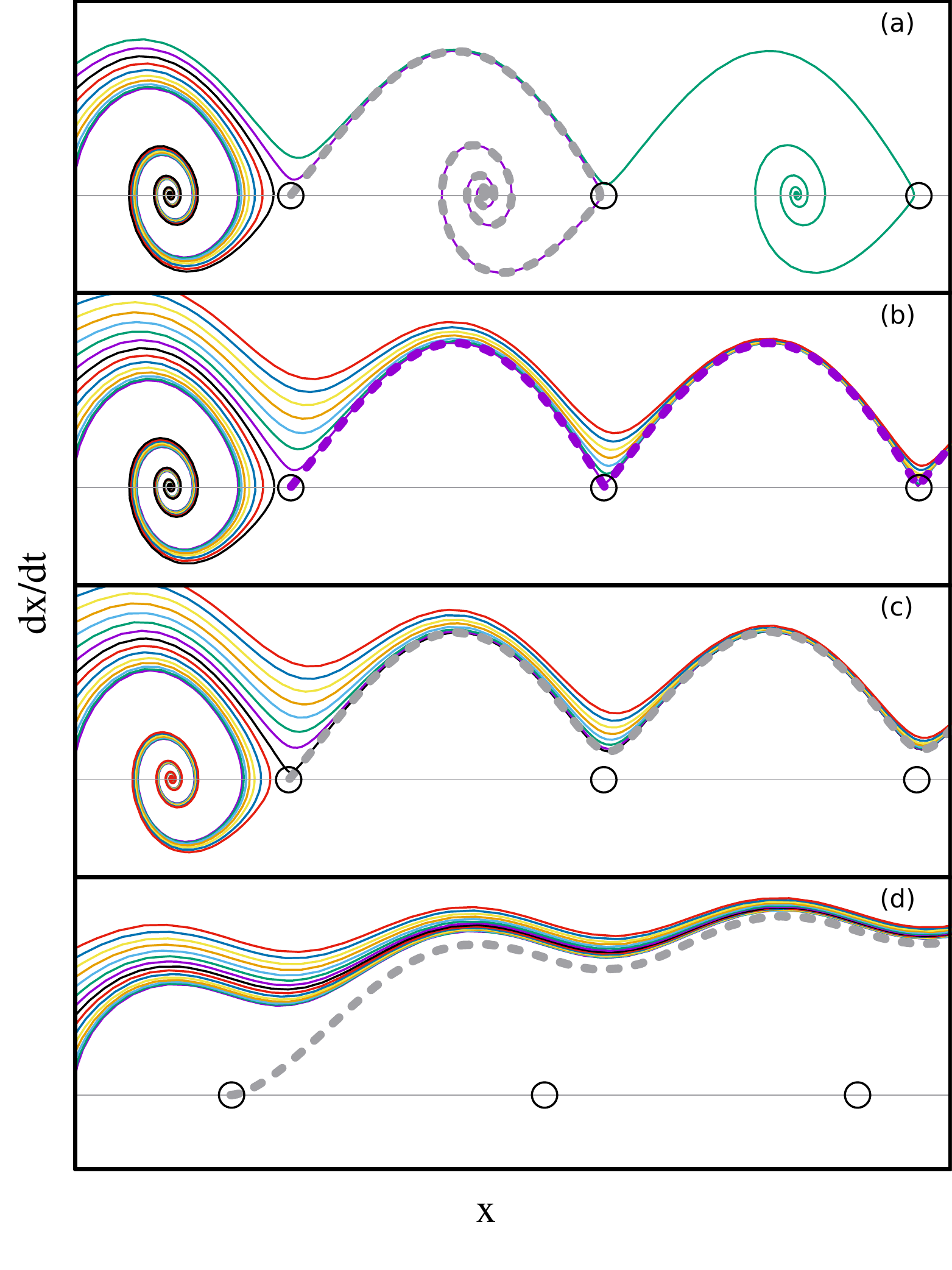}
\caption{
Phase-space trajectories obtained numerically from Eq.~\eqref{eq:mainequation}, using different
initial conditions, for the weakly damped case $\alpha<\alpha_c$.
In \textbf{(a)}, $f<f_c(\alpha)$: All initial conditions starting at the left margin of the plot
with different velocities are finally spirally attracted to one of the periodic images of the stable
fixed points of the tilted washboard potential, such that $V'(x)-f =0$ and $V''(x)>0$.
In \textbf{(b)}, $f=f_c(\alpha)$: Low initial velocity trajectories are trapped but large velocity
trajectories approach the homoclinic orbit (in purple dashed line).
In \textbf{(c)}, $f_c(\alpha)<f<f_c^0$: Trajectories are either trapped or acquire a finite
time-averaged velocity converging to the stable limit cycle.
In \textbf{(d)}, $f=f_c^0$: Stable fixed points disappear, and all trajectories are attracted to the
running periodic orbit with a finite average velocity.
With dashed lines we show trajectories starting with infinitesimally positive velocity at one of the
unstable fixed points $x^*$ of the tilted potential (open circles), such that $V'(x^*)-f=0$ and
$V''(x^*)<0$. The homoclinic orbit corresponds exactly to the dash line in (b).
}
\label{fig:phasespaceorbits}
\end{figure}

Central to our analysis is the computation of the critical trajectory $x^\star(t)$ that connects,
for $f_c(\alpha)<f<f_c^0$ and $\alpha<\alpha_c$, a local maximum of the tilted potential $V(x)-fx$
to the following one in an infinite time window.
In phase space, such trajectory, called the homoclinic orbit, acts as a separatrix:
It separates two domains of the initial conditions:
(i) those that lead to the limit cycle, which is characterized by a running periodic solution with
$\dot x(t)>0$,
and (ii) and those that end on a local minimum of the tilted  potential, with $\dot x(t)\to 0$ as
$t\to\infty$.
In Fig.~\ref{fig:phasespaceorbits}(c) we depict such two classes of trajectories.
In contrast, for $f<f_c(\alpha)$ all initial conditions are trapped in stable fixed points, while
for
$f\geq f_c(\alpha)$ only running solutions are stable,
as shown in Figs.~\ref{fig:phasespaceorbits}(a) and~\ref{fig:phasespaceorbits}(d), respectively.
An accurate numerical analysis of the homoclinic orbit is specially difficult near the triple point
$\alpha_c$ we are interested in, because the bistability range vanishes.
To make progress we will hence adopt a complementary approach by tackling its properties
analytically.

\section{A soluble tilted periodic potential}
\label{sec:def-pot-phi4}
\begin{figure}
\includegraphics[width=.9\columnwidth]{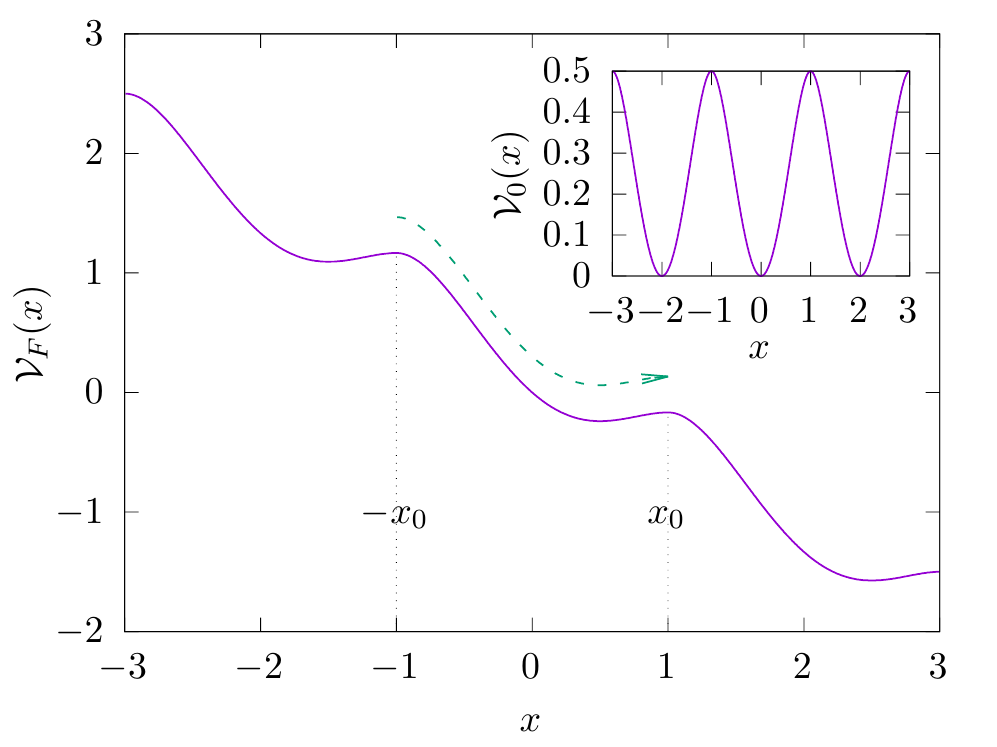}
\caption{
The nontilted (inset) and tilted (main) potentials
$\mathcal V_0(x)$ and $\mathcal V_F(x)$
defined by Eqs.~\eqref{eq:defV0}
and~\eqref{eq:defVf}, respectively (parameters are $\mu=g=1$ and $F=\frac 12 F_c^0$).
A critical trajectory $x^\star(t)$  of a massive particle, with the dynamics~\eqref{eq:dynVf}, is
represented with a green dashed line:
It joins the location $-x_0$ of a local maximum at time $t=-\infty$ to its periodic image $x_0$ at
time $t=\infty$.
}
\label{fig:V0Vf}
\end{figure}

In the definition~\eqref{eq:mainequation} of the model, we considered a generic pinning potential
$V(x)$ and we also took as an example potential a cosine function.
In this section, we study a special form of potential,
a periodically replicated quartic double-well potential.
It allows us to obtain the exact critical trajectory and to obtain the exponents of the critical
region of the homoclinic line close to the triple point.

Consider a potential
\begin{equation}
\label{eq:defV0}
\mathcal V_0(x) = -\frac g2 x^4 + \mu^2 x^2
\end{equation}
defined on $[-x_0,x_0]$ with $x_0=\frac \mu {\sqrt g}$.
Since it is even [i.e., $\mathcal V_0(-x_0)=\mathcal V_0(x_0)$]
we can make it periodic on $\mathbb R$.
We also denote $\mathcal V_0(x)$ (see Fig.~\ref{fig:V0Vf}) this periodic potential.

To \emph{model} the drive of  a particle living in such potential out of equilibrium, we tilt the
potential $\mathcal V_0(x)$ as
\begin{equation}
\label{eq:defVf}
\mathcal V_F(x) = \mathcal V_0(x) - \left( x-\frac{x^3}{3x_0^2}\right) F,
\end{equation}
where the constant $F$ represents the amplitude of a driving force.
The driving force can be made periodic in the same way as for $\mathcal V_0(x)$, in the sense that
the force $-\mathcal V'_F(x)$ corresponding to $\mathcal V_F(x)$  is periodic.
The resulting tilted potential $\mathcal V_F(x)$ is shown on Fig.~\ref{fig:V0Vf}.
The form~\eqref{eq:defVf} of the drive includes a cubic contribution on top of the usual linear
one,
which seems unnatural but allows us to find an exact expression of the critical trajectory
(the homoclinic orbit) at \emph{nonzero mass}.
The relation between the parameters $\{g,\mu,F\}$ of this effective model and
those of a physical one will be discussed in Sec.~\ref{sec:param-effect-model}.
We stress that the relation between the effective drive $F$ and the force $f$ of the original model
does not take a simple form on the whole range of forces of interest.
Note in passing that the form~\eqref{eq:defVf} of the drive ensures that the locations of the local
maxima of $\mathcal V_0(x)$ are unchanged: they remain in $\pm x_0$ as $F$ increases.

\subsection{An exact critical trajectory at nonzero mass}
\label{sec:an-exact-critical}
Let us consider the dissipative dynamics~\eqref{eq:mainequation} of a massive particle in the
potential $\mathcal V_F(x)$,
\begin{equation}
\label{eq:dynVf}
m \ddot x + \gamma \dot x = -\mathcal V'_F(x) \:.
\end{equation}
For values of the drive $F$ less than the critical drive
\begin{equation}
\label{eq:deffc}
F^0_c = 2 \frac{\mu^3}{\sqrt g} \:,
\end{equation}
we see that the tilted potential~\eqref{eq:defVf} presents local minima.
This implies that at zero mass the particle gets trapped into a local minimum and the average
velocity is zero.
At nonzero mass, if the initial position of the particle is close enough to a local minimum, then
the velocity at long times is also zero.
We now determine the condition on the mass allowing, in some range $F_c(m)<F<F^0_c$,
for the coexistence of another class of trajectories that converge to a limit cycle,
and hence possesses a nonzero average velocity $v$.

We find by direct computation that, for $F<F_c^0$, there exists a critical trajectory $x^\star(t)$
joining two local maxima of $\mathcal V_F(x)$ located in $-x_0$ and $x_0$, between times $t=-\infty$
and $t=\infty$ (see Fig.~\ref{fig:V0Vf}).
It takes the form
\begin{equation}
\label{eq:defxstar}
x^\star\!(t) = x_0 \tanh \frac t \tau
\qquad \text{with} \qquad
\tau = \frac{\sqrt{M_c(F)}}\mu
\:,
\end{equation}
provided the mass $m$ has the value
\begin{equation}
\label{eq:Mcf}
M_c(F) = \frac 14 \bigg(\frac{F_c^0}{F}\bigg)^2 \bigg(\frac\gamma\mu\bigg)^2 \:.
\end{equation}
The existence of such explicit solution for the critical trajectory is not immediate, because
in presence of dissipation ($\gamma>0$) and drive ($F>0$) the evolution equation~\eqref{eq:dynVf}
does not preserve
energy anymore and there is no conserved quantity along the trajectory.

\subsection{Critical mass and homoclinic bifurcation}
\label{sec:crit-mass-homocl}
The interpretation of the solution \eqref{eq:defxstar} and~\eqref{eq:Mcf} is that of a trajectory
$x^\star(t)$ that allows the particle to cross a barrier of potential (for $F<F_c^0$) by use of
inertia, provided its mass takes the precise value $M_c(F)$.
Such trajectory has  the threshold mass that allows it to store enough kinetic energy when going
downhill to precisely compensate for the friction-induced dissipation along its course.
We call such a trajectory an ``inertial critical trajectory.''
In mathematical terms, such a trajectory is a separatrix joining the unstable point $-x_0$ to its
periodic image $x_0$, allowing for a homoclinic bifurcation.

Physically, if we fix a drive $F<F_c^0$, and increase the mass $m$ starting from $m=0$ (keeping
every other parameters fixed), then the mass $M_c(F)$ is the first mass for which a trajectory
joining two local maxima of $\mathcal V_F(x)$ starts to exist.
Based on this, we now determine the critical mass $m_c$ of the dynamics that corresponds
(at fixed $\gamma$) to the critical damping $\alpha_c$.
We prove that the critical mass is equal to:
\begin{equation}
  \label{eq:mc}
  m_c = \frac 14  \bigg(\frac\gamma\mu\bigg)^2
  \:.
\end{equation}
For the demonstration of this relation, we denote $m^\star_c =\frac 14 \big(\frac\gamma\mu\big)^2$.
\begin{itemize}
\item \underline{Proof that $m_c\geq m_c^\star$}:
Consider a drive $F<F_c^0$.
For all masses $m<m_c^\star$, we see from~\eqref{eq:Mcf} that $m<m_c^\star<M_c(F)$, and hence there
exists no inertial critical trajectory.
We thus have proved that for all $m<m_c^\star$ the asymptotic velocity is 0.
Hence, as announced, $m_c\geq m_c^\star$.

\item \underline{Proof that $m_c\leq m_c^\star$}:
For a drive $F<F_c^0$, consider a mass $m>m_c^\star$.
We see from~\eqref{eq:Mcf} that there exists a domain of drive $[F_c(m),F_c^0]$ such that,
$\forall F\in[F_c(m),F_c^0]$, there is an inertial critical trajectory.
Since this is possible for all $m>m_c^\star$, this shows as announced that $m_c\leq m_c^\star$.
Note that the exact explicit expression of $F_c(m)$ is
\begin{equation}
\label{eq:fcm}
F_c(m) = \sqrt{\frac{m_c}{m}}\, F_c^0 \:.
\end{equation}
\end{itemize}
From the previous reasoning, we see that the expression~\eqref{eq:fcm}
is precisely that of the homoclinic bifurcation line of the model (on the range of mass $m>m_c$).
Translating the results~\eqref{eq:mc} and~\eqref{eq:fcm} from the variable $m$ (at fixed $\gamma$)
to the damping variable $\alpha=\gamma/\sqrt m$, one obtains the following expressions for the
critical damping $\alpha_c$ and the homoclinic line $F_c(\alpha)$:
\begin{align}
\label{eq:alphaceffective}
\alpha_c &= 2\mu \:,
\\
F_c(\alpha) &= \frac{\alpha}{2\mu} = \frac{\alpha}{\alpha_c}
\qquad \text{(for $\alpha\in[0,\alpha_c]$)} \:.
\end{align}
We emphasize that the relation between the effective drive $F$ and the physical force $f$ is not
direct:
As we now detail, these parameters are not scaling in the same way in the two regimes of interest
(the low-damping limit and the vicinity of the triple point).
In particular, if the homoclinic line $F_c(\alpha)$ is linear in $\alpha$ when $\alpha$ approaches
$\alpha_c$ from below, then this will not be the case for $f_c(\alpha)$.

\begin{figure}
\begin{center}
\includegraphics[width=.9\columnwidth]{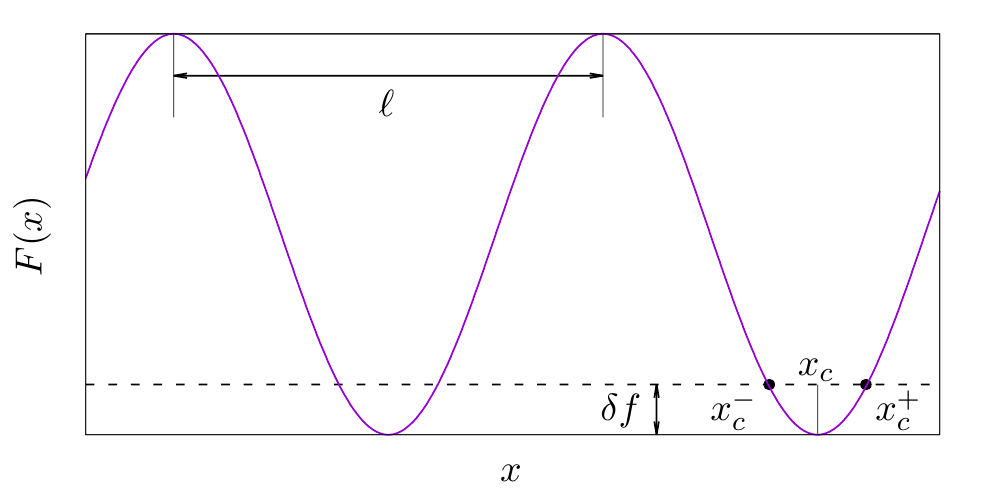}
\end{center}
\caption{
Landscape of force $F(x) = -V'(x)+f$ at external force~$f$ close to depinning ($f=f_c^0-\deltaf $
with $\deltaf>0$), presenting the critical points $x_c^\pm=x_c\pm\sqrt{\deltaf/\kappa}\,$.}
\label{fig:Fxcrit}
\end{figure}

\subsection{Parameters of the effective model close to the depinning point of a
physical model}
\label{sec:param-effect-model}
We now come back to the model of Eq.~\eqref{eq:mainequation} with a generic periodic potential
$V(x)$ of spatial period $\ell$, and we assume that it is monotonous between its unique minimum and
the closest maxima  (see Fig.~\ref{fig:Fxcrit}).
A concrete example is given by the cosine potential
\begin{equation}
\label{eq:defV0sin}
V_{\cos}(x) = \mathfrak f_0 \ell\, \frac{\cos\frac{2\pi x}{\ell}} {2\pi}\:.
\end{equation}
The corresponding force $F(x)$, to which the particle is subjected,
includes an external constant force $f$ with $F(x)=-V'(x)+f$.
For the cosine potential, this gives:
\begin{equation}
\label{eq:defF0sin}
F_{\cos}(x) = \mathfrak f_0  \sin\frac{2\pi x}{\ell} + f\:.
\end{equation}
Along a period, the contribution $-V'(x)$ to the force presents a minimum in $x_c$. The
corresponding critical force (for the zero-mass depinning) is
\begin{equation}
\label{eq:fc0depinning}
f_c^0 = -V'(x_c) \:.
\end{equation}
For an external force $f<f_c^0$ close to the $m=0$ depinning force, denoting $f=f^0_c-\deltaf$
(with $\deltaf>0$ and $\deltaf\ll f^0_c$), the dynamics presents two critical points $x_c^\pm$
represented on Fig.~\ref{fig:Fxcrit}.
Writing that, close to $x_c$,
\begin{equation}
\label{eq:devFxc}
F(x_c+\delta x) = -\deltaf + \kappa \,\delta x^2 + \ldots
\end{equation}
we find
\begin{equation}
\label{eq:xcpm}
x_c^\pm = x_c \pm \sqrt{\frac{\deltaf}{\kappa}} + \ldots
\end{equation}
Note that for the cosine potential~\eqref{eq:defV0sin} one has
$f^0_c=\mathfrak f_0$, $x_c=\frac 34 \ell$, $\kappa=2\pi^2 \mathfrak f_0/\ell^2$.

\smallskip
We are now in position to determine the values of the parameters $\{g,\mu,F\}$ of the effective
potential~\eqref{eq:defV0} and~\eqref{eq:defVf} that describe the criticality of interest.
We fix
\begin{align}
\label{eq:resg}
g = 4 \frac{\mu^2}{\ell^2} \:,
\end{align}
ensuring that the spatial period of the potential $\mathcal V_0(x)$ is equal to $\ell$.
Then, shifting $x$ so that $-\mathcal V'_F(x^+_c)=0$, we impose the following conditions:
\begin{equation}
\label{eq:conditionsVprimeF}
-\mathcal V'_F(x^-_c)=0
\qquad\text{and}\qquad
-\mathcal V'_F(x_c)= -\deltaf
\:,
\end{equation}
which ensure that force $-\mathcal V'_F(x)$ of the effective model presents the critical properties
of $F(x)$ described on Fig.~\ref{fig:Fxcrit}.
Solving for $\mu$ and $F$, we thus find, for small $\deltaf$:
\begin{align}
\label{eq:Fofparams}
\mu &= \sqrt{\frac{\kappa \ell}{8}} + \sqrt{\frac{\deltaf }{32 \ell}} \ + \ O(\deltaf) \:,
\\
\label{eq:muofparams}
F &= \frac{\kappa  \ell ^2}{8}-\frac{3}{8} \sqrt{\kappa\,\deltaf }\, \ell \ + \ O(\deltaf) \:.
\end{align}
Note that, in these expressions, we should keep {both} contributions of order
$O(\sqrt{\vphantom{|}\deltaf})$ to describe correctly the critical scaling.

\subsection{Scaling regime \textnormal{$m\gtrsim m_c$} and
\textnormal{$f_c(m)\lesssim f_c^0$}}
\label{sec:scal-regime-mgtrs}
We have seen in Sec.~\ref{sec:crit-mass-homocl} that,
for masses below the critical mass $m_c$ given by~\eqref{eq:mc},
the bistable regime drive is $[F_c(m),F_c^0]$ with $F_c^0$ given by~\eqref{eq:deffc} and $F_c(m)$
by~\eqref{eq:fcm}.
Using the correspondence~\eqref{eq:Fofparams} and~\eqref{eq:muofparams} between the parameters of
the effective model and those of the physical model, we find that
\begin{equation}
\label{eq:scalingregimeclosetoFc0}
F_c^0 -F = \frac 12 \ell \sqrt{\kappa \,\deltaf} \ +\ O(\deltaf) \:.
\end{equation}
This relation implies an important scaling:
Close to the depinning point $f_c^0$ of the physical model ($f=f_c^0-\deltaf$),
the drive $F$ of the effective model scales as $F_c^0-F\sim\sqrt{\deltaf}$ and not as $\sim\deltaf$,
contrarily to what we could have naively expected.

Then, considering a mass $m$ slightly above the critical mass $m_c$,
\begin{equation}
\label{eq:mdeltam}
m = m_c+\deltam \qquad\text{with}\qquad \deltam \ll m_c \:,
\end{equation}
we see from~\eqref{eq:fcm} that the size of the bistable regime is determined by
\begin{equation}
\label{eq:sizebistablewindow}
F_c^0-F_c(m)
= \bigg(1-\sqrt{\frac{m_c}{m_c+\deltam}}\bigg) F_c^0
= \frac 12 F_c^0\,\frac{\delta m}{m_c} + O(\deltam^2).
\end{equation}
Thus, we see from~\eqref{eq:scalingregimeclosetoFc0} that, for the physical force $f$, the size
$\deltaf = f_c^0-f_c(m)$ of the bistable regime is governed by the scaling
$f_c^0-f_c(m) \propto (m-m_c)^2$;
more precisely:
\begin{equation}
\label{eq:scalingregimeclosetofc0exact}
f_c^0- f_c(m) = \frac{\ell^4\kappa^3}{256\,\gamma^4 }\big(m-m_c\big)^2
\quad \text{for } m\to m_c^+\:.
\end{equation}
For the damping coefficient, we get
\begin{equation}
\label{eq:scalingregimeclosetofc0exactalpha}
f_c^0- f_c(\alpha) =  \kappa^3 \left[\frac{\ell}{64\,\mu^3 }\right]^2 \big(\alpha_c-\alpha\big)^2
\quad \text{for } \alpha\to \alpha_c^-\:.
\end{equation}

\smallskip
For the cosine potential~\eqref{eq:defV0sin} we have $\kappa=2\pi^2 \mathfrak f_0 / \ell^2$ and
thus
\begin{equation}
\label{eq:scalingregimeclosetofc0exactcos}
f_c^0- f_c(m) =  \frac{\pi^6}{32}\, \frac{\mathfrak f_0^3}{\ell^2\,\gamma^4 }\big(m-m_c\big)^2 \:,
\end{equation}
where ${\pi^6}/{32} \approx 30.0\,$.

\subsection{Effective description on the full range of forces
\textnormal{$[0,f_c^0]$} for the tilted cosine potential}
\label{sec:effect-descr-full}
For the cosine potential~\eqref{eq:defV0sin} the zeros $x_c^\pm$ of the corresponding
force~\eqref{eq:defF0sin} are given (see Fig.~\ref{fig:Fxcrit}) by
\begin{equation}
\label{eq:xcpxcmcos}
x_c^- = \frac{\ell }{2} + \frac{\ell  \arcsin \frac{f}{\mathfrak{f}_0}}{2 \pi }
\qquad\text{and}\qquad
x_c^+ = \ell -\frac{\ell  \arcsin \frac{f}{{\mathfrak f_0}}}{2 \pi }.
\end{equation}
Performing the same program as previously in order to find the parameters of the effective tilted
potential $\mathcal V_F$, we impose
\begin{equation}
\label{eq:conditionsVprimeFbis}
\mathcal V'_F(x^-_c)=0
\qquad\text{and}\qquad
-\mathcal V'_F(x_c)= f-f^0_c
\:,
\end{equation}
where $f_c^0=\mathfrak f_ 0$.
We shift the $x$ coordinate to ensure $\mathcal V'_F(x^+_c)=0$
(before the shift one has $x_c = \frac 34 \ell$).
We find
\begin{align}
\label{eq:Fmucosgeneric1}
F
&= \frac{4 \pi ^2 \left(\mathfrak f_0-f\right) \arcsin\frac{f}{\mathfrak f_0}}
  {\left[3 \pi+2 \arcsin \frac{f}{\mathfrak f_0} \right]
    \left[\arccos\frac{f}{\mathfrak f_0}\right]{}^2}
\\
\label{eq:Fmucosgeneric2}
\mu
&= \frac{\pi ^{3/2}}{\arccos\frac{f}{\mathfrak f_0}
  \sqrt{\frac{\ell}{\mathfrak f_0-f} \left(\frac{3 \pi }{2}+\arcsin\frac{f}{\mathfrak f_0}\right)}}
\:,
\end{align}
where $g$ is again given by~\eqref{eq:resg}.

From Eqs.~\eqref{eq:deffc}, \eqref{eq:mc}, and~\eqref{eq:fcm},
we find that the corresponding equation for the pinned or bistable homoclinic critical line
$f_c(m)$,
\begin{equation}
\label{eq:relationfm}
\frac{4 \sqrt{\vphantom{|^|}\mathfrak f_0-f_c(m)} \arcsin\frac{f_c(m)}{\mathfrak f_0}}
  {\gamma \arccos \frac{f_c(m)}{\mathfrak f_0}\ \sqrt{ \ell \left(2-\frac{1}{\pi}
    \arccos\frac{f_c(m)}{\mathfrak f_0}\right)}}
= \frac{1}{\sqrt{m}}
\:.
\end{equation}
We easily check that expanding this relation for $m$ close to $m_c$ and $f$ close to
$f_c^0=\mathfrak f_0$ (i.e., $f=\mathfrak f_0-\deltaf$, $\deltaf\ll \mathfrak f_0$) we recover the
result~\eqref{eq:scalingregimeclosetofc0exactcos}, which describes the behavior of the homoclinic
line close to the triple point.

In the other asymptotics, for small $f$ (which corresponds to large mass along the homoclinic line),
we find
\begin{equation}
\label{eq:Fmusmallf}
F = \frac{16}{3\pi} f
\quad\text{and}\quad
\mu = 2 \sqrt{\frac{2}{3}\frac{\mathfrak f_0}{\ell}}
\qquad\text{for}\;\:
f\ll\mathfrak f_0
\:.
\end{equation}
In this regime, the effective drive $F$ is proportional to the tilt force $f$,
as physically expected.
From Eq.~\eqref{eq:mc}, the effective critical mass in that regime is found to be
\begin{equation}
\label{eq:mcsmallf}
m_c = \frac{3}{32} \frac{\ell \gamma^2}{\mathfrak f_0} \:,
\end{equation}
and we find from~\eqref{eq:fcm} that the critical equation for the line
between the pinned and the bistable regime is
\begin{equation}
\label{eq:fmseparatrixsmallf}
f_c(m)\sim \frac{\pi}{8} \sqrt{\frac{3\mathfrak f_0\ell}{2m}}\,\gamma
\qquad\text{for}\;\:
m\gg m_c
\:.
\end{equation}
We should beware that the result of Eq.~\eqref{eq:mcsmallf} on the location of the triple point is
only indicative since it results from a computation done in the regime of small forces along the
homoclinic line (that is, far from the triple point).
Translating the result~\eqref{eq:fmseparatrixsmallf} to the damping variable $\alpha$, one finds
\begin{equation}
\label{eq:fmseparatrixsmallfalpha}
f_c(\alpha)\sim \frac{\pi}{8} \sqrt{\frac{3\mathfrak f_0\ell}{2}}\,\alpha
\qquad\text{for}\;\:
\alpha\ll\alpha_c
\:.
\end{equation}
We recover the expected scaling $f_c(\alpha)\propto\alpha$ of the large-damping limit.
For the parameters $\mathfrak f_0=1$ and $\ell=2\pi$ corresponding to the cosine potential of
Guckenheimer and Holmes~\cite{guckenheimer1983}, the prefactor becomes $\frac 18 \sqrt{3}
\pi^{3/2}\simeq 1.21$ which is not very far from the exact prefactor $4/\pi\simeq 1.27$.
This result validates our approach based on an approximate ``periodicized $\varphi^4$'' potential.
A derivation of the exact prefactor $4/\pi$ and of its generalization for an arbitrary potential is
done in Sec.~\ref{sec:gener-scal-theory}.

\subsection{The critical mass and its asymptotic behaviors}
\label{sec:critical-mass-its}
We note that in the regime of forces $f$ close to the triple point, critical mass coming from
Eqs.~\eqref{eq:mc} and~\eqref{eq:Fofparams} is
$\frac{1}{\pi^2} \frac{\ell \gamma^2}{\mathfrak f_0}$,
while we obtained a different numerical prefactor in~\eqref{eq:mcsmallf}.
The reason behind this mismatch is that the approach we follow consists in approximating the tilted
potential $V(x)-fx$ by the effective one, $\mathcal V_F(x)$, and that the effective parameters
$\mu$, $g$ and $F$ depend on $f$ in a nontrivial way, determined by the homoclinic line force
$f_c(m)$.
The first result is derived for the asymptotics $f_c(m)-f_c^0\ll f_c^0$ and the second one for
$f_c(m)\ll f_c^0$.
This means that our approach predicts
\begin{equation}
\label{eq:mcregimes}
m_c \approx
  \left\{
    \begin{aligned}
      \frac{3}{32} \frac{\ell \gamma^2}{\mathfrak f_0}
        & \qquad \text{for }\quad \frac{\ell \gamma^2}{\mathfrak f_0}\to 0
      \\[2mm]
      \frac{1}{\pi^2} \frac{\ell \gamma^2}{\mathfrak f_0}
        & \qquad \text{for }\quad \frac{\ell \gamma^2}{\mathfrak f_0}\to\infty
      \:,
    \end{aligned}
  \right.
\end{equation}
corresponding respectively to a regime where dissipation is low (compared to the potential barriers)
and a regime where dissipation is higher and close to the maximal one allowing for a limit cycle at
$f<f_c^0$.
We note that the numerical prefactors in both cases of~\eqref{eq:mcregimes} are rather close:
This means in the intermediate regime where $\frac{\ell \gamma^2}{\mathfrak f_0}$ takes a finite
value, the critical mass scales as in~\eqref{eq:mcregimes} with a prefactor mildly depending on
$\frac{\ell \gamma^2}{\mathfrak f_0}$.

\subsection{Mapping for general normal forms: $\Upsilon \neq 2$}
\label{sec:genupsilon}
Considering a tilted force that is expanded close to its critical point as
$f-V'(x) \approx (f-f_c^0)+k|x-x^*|^\Upsilon$
instead of~\eqref{eq:devFxc}, one simply replaces $\sqrt{\deltaf}$ by $\deltaf^{1/\Upsilon}$ in
Eq.~\eqref{eq:xcpm}.
This implies that the same substitution has to be done in Eqs.~\eqref{eq:Fofparams}
and~\eqref{eq:muofparams}.
Instead of~\eqref{eq:scalingregimeclosetoFc0}, one obtains now
\begin{equation}
\label{eq:Fc0FUps}
F_c^0-F \sim \deltaf^{1/\Upsilon}\quad\text{for } \deltaf\to 0\:.
\end{equation}
Then, Eqs.~\eqref{eq:scalingregimeclosetofc0exact} and~\eqref{eq:scalingregimeclosetofc0exactalpha}
become
\begin{align}
\label{eq:fcmUps}
f_c^0-f_c(m) & \sim (m-m_c)^\Upsilon \:,
\\
\label{eq:fcmUpsalpha}
f_c^0-f_c(\alpha) & \sim (\alpha_c-\alpha)^\Upsilon
\:,
\end{align}
and are valid in the vicinity of the triple point ($m\to m_c^+$, i.e., $\alpha\to \alpha_c^-$).
The large damping regime, which is more universal, is described at the end of
Sec.~\ref{sec:pert-at-small}.

\section{Far from the triple point: the large-damping regime}
\label{sec:gener-scal-theory}

\subsection{Settings}
\label{sec:settings_largeM}
To understand the scaling properties of the large-damping regime, we consider that the potential
$V(x)$ is described by two physical parameters, its period $\ell$ and its amplitude $V_0$:
\begin{equation}
\label{eq:defVscaling}
V(x) = V_0\,\hat V(x/\ell) \:,
\end{equation}
where one assumes that the rescaled potential $\hat V(\hat x)$ is independent of $V_0$ and $\ell$.
At zero friction ($\gamma=0$) and zero external force ($f=0$) the equation of
motion~\eqref{eq:mainequation} for the position of the
particle---whose coordinate is denoted by $x_0(t)$---becomes
\begin{equation}
\label{eq:motionzeromassscaling}
m \ddot x_0  = -V'(x_0)
\end{equation}
which, on the rescaling
\begin{equation}
\label{eq:defxhattauscaling}
x_0(t) = \ell\,\hat x_0(t/\tau) \qquad\text{with }\quad \tau=\sqrt{\frac{m\ell^2}{V_0}}
\end{equation}
becomes
\begin{equation}
\label{eq:eqmxhatzerofriction}
\ddot {\hat x}_0 = -\hat V'(\hat x_0) \:.
\end{equation}
Since the dynamics is conservative, its description is rather explicit.
Its lowest energy solution, which starts from a local maximum of
$\hat V(\hat x)$---located in $\hat x=0$, without loss of generality---with an infinitesimally
small velocity at time $t=-\infty$ and arrives at the next maximum in
$\hat x=1$ at time $t=+\infty$, is the solution of the differential equation
\begin{equation}
\label{eq:eqmxhatenergy}
\frac 12 \left[ \dot {\hat x}_0\left(\hat t\right)\right]^2
= \hat V(0)-\hat V\left[\hat x_0\left(\hat t\right)\right]
\:.
\end{equation}
We denote its solution by $\hat x_0^\star(\hat t)$; it is independent of the physical parameters
$V_0$ and $\ell$ and is given by
\begin{align}
\label{eq:relthatshat0sol}
\dd\hat t
= \frac{\dd\hat x_0^\star}{\sqrt{2\big[\hat V(0)-\hat V(\hat x_0^\star) \big]}}
\:,\quad\text{i.e.}
\\
\label{eq:relthatshat0solb}
\hat t
= \int_0^{\hat x_0^\star(\hat t)}\frac{\dd\hat x_0}{\sqrt{2\big[\hat V(0)-\hat V(\hat x_0) \big]}}
\:.
\end{align}

\subsection{Perturbation at small friction and small drive}
\label{sec:pert-at-small}
To describe the small-dissipation asymptotics of the homoclinic curve,
we now drive the system by applying a small force $f$ and adding a small dissipation  $\gamma$ that
ensures the energy remains finite.
The motion is described by Eq.~\eqref{eq:mainequation}.
For a given friction $\gamma$ and a mass $m$, we are looking for the critical value of the force
$f_c(m)$ above which a bistable regime is possible.
This homoclinic line $f_c(m)$ is determined by the existence of a critical solution $x^\star(t)$
to~\eqref{eq:mainequation} with the boundary conditions
\begin{align}
\label{eq:bccritx}
x(-\infty) = x_c
& \qquad\text{;}\qquad
x(\infty) = x_c+\ell
\\
\label{eq:bccritx2}
\dot x(-\infty) = 0^+
& \qquad\text{;}\qquad
\dot x(\infty) = 0
\,,
\end{align}
where $x_c$ is the location of the maximum of $V(x)-f x$ (satisfying $x_c\to 0$ as $f\to 0$).
On the rescaling~\eqref{eq:defxhattauscaling} one finds
\begin{equation}
\label{eq:eqmxhat}
\ddot {\hat x} = -\hat V'(\hat x)
\underbrace{-\frac{\gamma \tau}{ m} \,\dot {\hat x} + \hat f}_{\text{``small''}}
\qquad\text{where}\quad
\hat f = \frac{f}{V_0/\ell}
\,.
\end{equation}
We consider the rescaled critical solution $\hat{x}^\star(t)$.
Multiplying~\eqref{eq:eqmxhat} by $\dot{\hat{x}}^\star(t)$, integrating between $t=-\infty$ and
$t=+\infty$ and using the boundary conditions~\eqref{eq:bccritx} and~\eqref{eq:bccritx2}, together
with
$[\hat V]_{\hat x_c}^{\hat x_c+1}=0$ (by periodicity), one finds that necessarily
\begin{equation}
\label{eq:condxhat}
\frac{\gamma \tau}{ m} \int_{-\infty}^{\infty} \dd \hat t\; \big(\dot{\hat x}^\star\big)^2 = \hat f
\,.
\end{equation}
This relation is true in general for the inertial critical trajectory (for any $f$ and $\gamma$ at
the critical value of the mass) and expresses the fact that the energy dissipated along this
trajectory matches exactly the potential loss $\hat f$ on one period.
If one is able to determine the expression of the critical trajectory $\hat x^\star(t)$ then
Eq.~\eqref{eq:condxhat} allows one to obtain the expression of $f_c(m)$.
However this is not possible in general.
In our asymptotics of interest ($f\to 0$, $\gamma\to 0$), since both sides are small one can replace
$\dot{\hat x}^\star$ by the zero-friction zero-force solution $\dot{\hat x}_0^\star$.
This is the essence of Melnikov's formalism~\cite{melnikov1963} but written in a theoretical
physicist's manner.
Then, using~\eqref{eq:relthatshat0sol} to convert the time integral into a spatial integral
together
with the expression~\eqref{eq:defxhattauscaling} of the characteristic time $\tau$, one finds
\begin{equation}
\label{eq:relationmc0}
\frac{\gamma \ell}{\sqrt{m\, V_0}} \int_0^1 \dd \hat x\: \sqrt{2\big[\hat V(0)-\hat V(\hat x) \big]}
= \hat f
\,.
\end{equation}
This relation gives the criterion relating $f_c(m)$ and in the small damping limit:
\begin{align}
\label{eq:relationmc}
f_c(m)
&= \mathcal N \,\frac{\gamma\, \sqrt{\vphantom{|^1}V_0}}{\sqrt{m}}
\qquad \text{for }
\frac{\gamma}{\sqrt{m}}\to 0
\\[2mm]
\label{eq:relationmcdefN}
\mathcal N
&= \int_0^1 \dd \hat x\: \sqrt{2\big[\hat V(0)-\hat V(\hat x) \big]}
\,.
\end{align}
Here the prefactor $\mathcal N$ is a numerical constant.
In terms of the damping coefficient, the homoclinic asymptotics writes
$f_c(\alpha)\sim \mathcal N \sqrt{V_0}\,\alpha$ for $\alpha\to 0$.
For the cosine potential, one has $\hat V(\hat x) = \cos 2\pi \hat x$ and $\mathcal N=\frac 4\pi$:
One recovers the result of Levi \emph{et al.}~\cite{levi1978dynamics} and
Guckenheimer and Holmes~\cite{guckenheimer1983}.
We note that the predictions of Eqs.~\eqref{eq:relationmc} and~\eqref{eq:relationmcdefN} are
independent on the regularity properties of the potential, so that they should be valid even in
presence of cusps.
We validate this prediction numerically in Sec.~\ref{sec:UpsilonPotential}.

\section{Numerical Validation}
\label{sec:0d_numerics}

\subsection{The case of the cosine potential}
\label{sec:case-cosine-potent}
In this section, we will validate our analytical results by using a numerical integration of
Eq.~\eqref{eq:mainequation} with the cosine potential of Eq.~\eqref{eq:defV0sin}.
Without loss of generality, we choose $\gamma=1$, $\mathfrak f_0=1$ and $\ell=2\pi$,
to simplify the notation.

\begin{figure}[tb]
\centering
\includegraphics[width=.9\columnwidth]{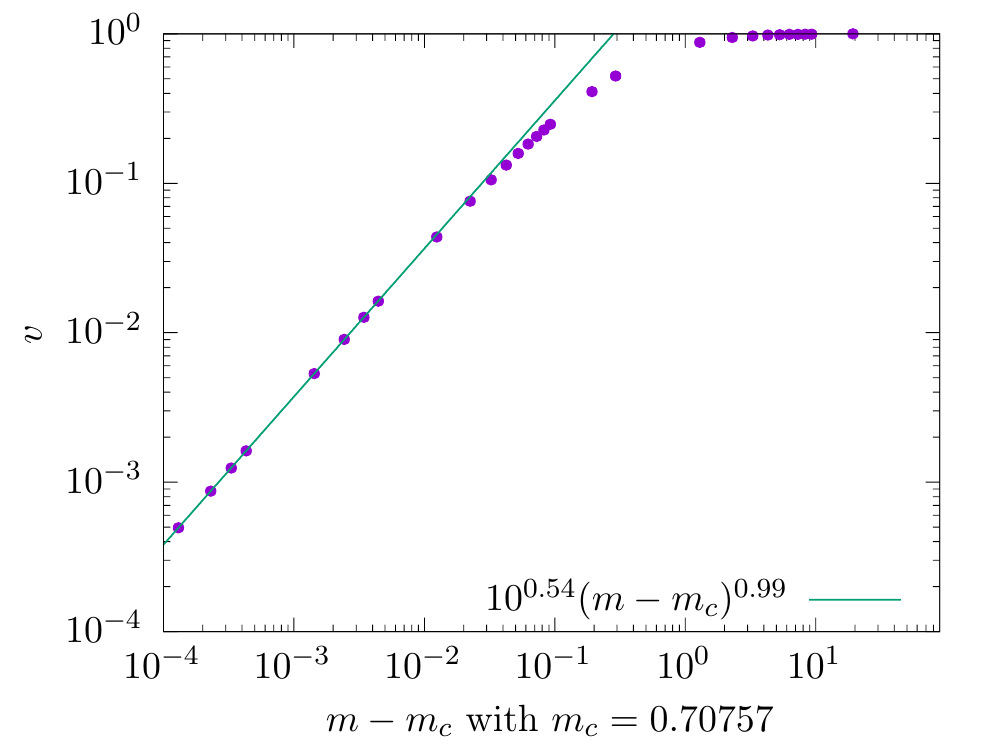}
\caption{
Steady-state time-averaged velocity as a function of the mass above the critical mass, $m-m_c$.
With a continuous green line we show a fitted power law,
where the exponent and the critical mass are fitting parameters.
Then, we find $m_c=0.70757 \pm 0.00002$.
}
\label{fig:v_vs_m}
\end{figure}

As we showed in Fig.~\ref{fig:v_vs_f}, we obtained the steady-state time-averaged velocity of the
particle, as a function of the driving force, and for different mass values.
This velocity seen as a function of the mass, when the driving force is equal to the critical force
of the massless particle, behaves as a power law above the critical mass, which, naively, can be
described as
\begin{equation}
\label{eq:power_law_m-mc}
v(f\equiv f_c^0)\sim(m-m_c)^{\beta\delta}.
\end{equation}
This allows us to determine the critical mass with great precision.
Leaving $m_c$, $\beta\delta$, and a prefactor as fitting parameters, we find
\begin{equation}
\label{eq:mc_value}
m_c = 0.70757 \pm 0.00002,
\end{equation}
and $\beta\delta = 0.99 \pm 0.01$, as we show on Fig.~\ref{fig:v_vs_m}.

Using the correspondence between $m$ and $\alpha=\gamma/\sqrt{m}$, we estimate from our numerical
results that the critical damping is $\alpha_c = 1.18882 \pm 0.00002$,
which seems to be compatible with the results from Ref.~\cite{levi1978dynamics},
even if there is no explicit numerical value given by the authors.
However, when comparing our estimations from Sec.~\ref{sec:critical-mass-its}
with the fitted value given by Eq.~\eqref{eq:mc_value}, we only find a mild agreement.
From Eq.~\eqref{eq:mcregimes}, we predicted a critical mass in the range $[0.589\dots,0.637\dots]$.
This mismatch comes from the fact that we are using a crude ``periodicized $\varphi^4$ potential''
approximation of the cosine potential, close to the triple  point of the phase diagram.
The numerical value of $m_c$ should depend on the exact properties of the potential on that point,
thus it is not a universal property of the model and
then it can only be partially estimated using that kind of approximation.
The order of magnitude is correct, and, to go further, one would need to find the critical
trajectory of a better approximation of the tilted potential.

\begin{figure}[tb]
\centering
\includegraphics[width=.9\columnwidth]{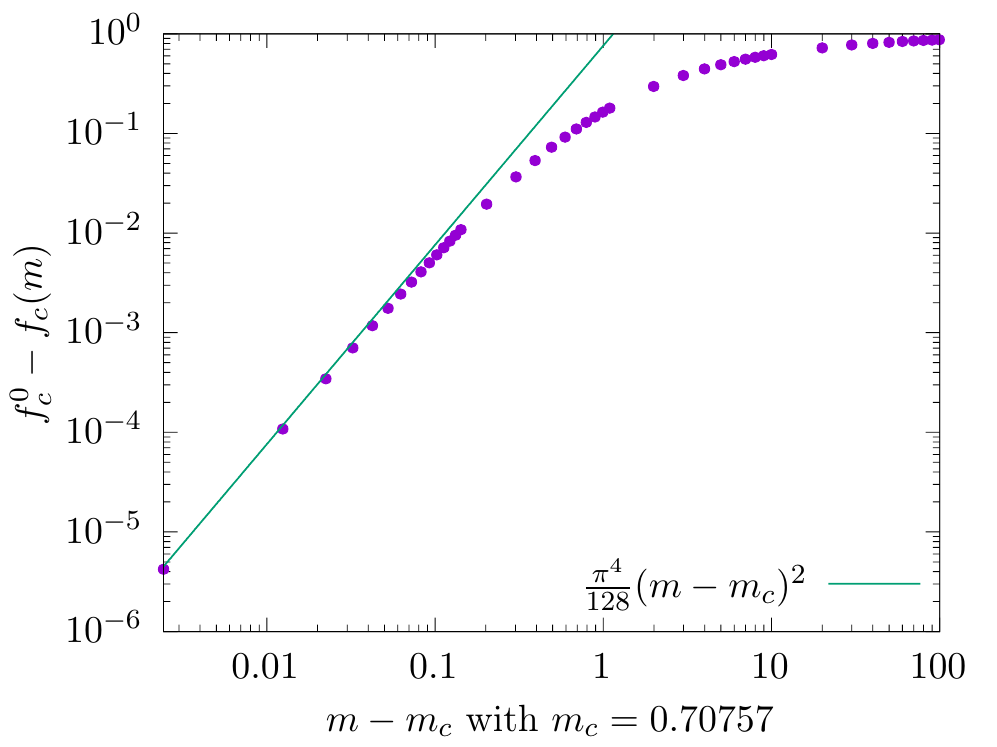}
\caption{
For the potential $V(x)=-\cos(x)$,
as the mass of the particle gets close to the critical mass from above,
the difference between the critical forces tends to zero as given
by Eq.~\eqref{eq:scalingregimeclosetofc0exactcos},
indicated by a continuous green line.
Here, we employ a numerical simulation using
$\ell \equiv 2\pi$, $\gamma\equiv1$ and $\mathfrak{f}_0 \equiv1$.
}
\label{fig:fc_vs_m}
\end{figure}

Besides the critical mass, we also estimate $\beta\delta$ to be near $1$.
Since $\beta=1/2$ for this model, this means $\delta=2$, which is in good agreement with the
prediction from Eq.~\eqref{eq:scalingregimeclosetofc0exact}.
Moreover, from Eq.~\eqref{eq:scalingregimeclosetofc0exactcos} and for our choice of potential
parameters, we find $f_c^0-f_c(m)=\frac{\pi^4}{128} (m-m_c)^2 $ close to the triple point.
Comparing this to our numerical data, using the previously estimated $m_c$, we find an excellent
agreement down to the prefactor, as can be seen in Fig.~\ref{fig:fc_vs_m}.
This could come as a surprise, as the prediction for the critical mass
was only approximate.
On the one hand, the  exponent $\delta$ predicted by Eq.~\eqref{eq:scalingregimeclosetofc0exact}
should be a universal exponent, hence more robust in the approximations made.
On the other hand, the prefactor on Eq.~\eqref{eq:scalingregimeclosetofc0exactcos} was specialized
for the cosine potential, but even with the other approximations made and the fact that the
prefactor should not be a universal property, the agreement is remarkably good.

\begin{figure}[tb]
\centering
\includegraphics[width=.9\columnwidth]{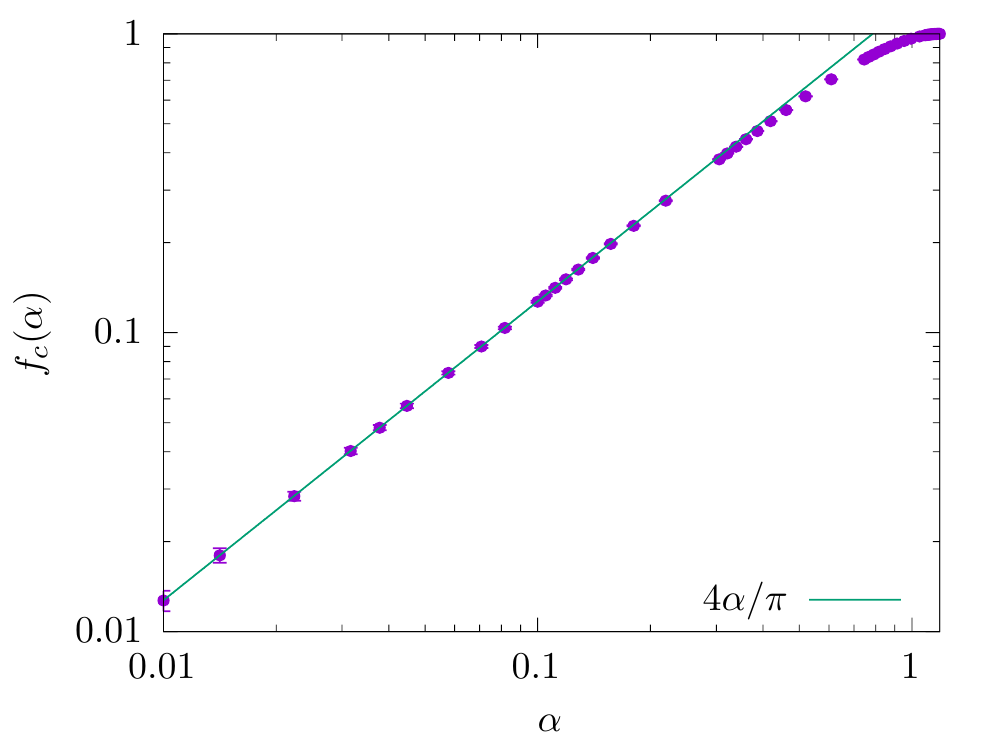}
\caption{
Critical force as a function of the damping parameter $\alpha$, defined as $\alpha=\gamma/\sqrt{m}$.
The expected behavior, from Eq.~\eqref{eq:fc_en_alpha_to_0},
is correct on the limit $\alpha \ll \alpha_c$.
}
\label{fig:fc_vs_alpha}
\end{figure}

Last, when considering the limit $m\to\infty$, we find the known behavior
from~ Ref.\cite[Eq.~(4.6.24)]{guckenheimer1983}
which, a function of the damping parameter, as $\alpha\to0$, is expressed as:
\begin{equation}
\label{eq:fc_en_alpha_to_0}
f_c(\alpha) = 4\alpha/\pi
\end{equation}
(see Sec.~\ref{sec:gener-scal-theory} for a simple demonstration).
On Fig.~\ref{fig:fc_vs_alpha}, we show that this analytical prediction
is in good agreement with our numerical results.

\subsection{Generic normal forms}
\label{sec:UpsilonPotential}
Here, we consider the more general periodic pinning force
\begin{align}
\label{eq:genericpinningforce}
-V'(x)
= \frac{\left(\sqrt{\pi } 2^{-\frac{\Upsilon }{2}} \Gamma \left(\frac{\Upsilon }{2}+1\right)\right)
  (1-\cos (x))^{\Upsilon /2}}{\Gamma \left(\frac{\Upsilon +1}{2}\right)}-1
\end{align}
with $\Upsilon>1$, where we have set the constant factors such that the upper critical force is
$f_c^0 \equiv \max_x[V'(x)]=V'(x_c)=1$
with the marginal points fixed at $x_c=2\pi n$ with $n$ integer.
Note that for $\Upsilon=2$, Eq.~\eqref{eq:genericpinningforce} reduces to the pinning force
$-V'(u)=-\cos(u)$ or simple washboard potential $V(u)= V(0) + \sin(x)$.
In the $\alpha \to \infty$ overdamped limit this periodic pinning force gives rise to the normal
form $\dot{x} \approx (f-f_c^0)+ |x-x_c|^\Upsilon$ displaying the depinning transition
$v\sim (f-f_c^0)^\beta$ with $\beta=1-1/\Upsilon$.
Note that this result remains valid for $\alpha>\alpha_c$ finite damping and in general
\begin{equation}
v\sim B(\Upsilon,\alpha) (f-f_c^0)^{1-1/\Upsilon}.
\end{equation}
Therefore inertia does not change the critical behavior of the velocity for all the infinite-period
bifurcation line $\alpha>\alpha_c$ although the prefactor may be affected.
This result can be appreciated in Fig.~\ref{fig:v_vs_f_diffm},
below the critical mass, i.e., $0\leq m\leq 0.7$.

To determine the critical mass $m_c$ and the behavior of $f_c(m)$
(particularly near the triple point $m_c$ and in the $m \gg m_c$ limit), instead of integrating
Eq.~\eqref{eq:mainequation} in time we have solved the equation
\begin{equation}
\label{eq:phasespaceeqmotion}
\frac{dK}{dx} = -\gamma \sqrt{\frac{2K}{m}} -V'(x) + f,
\end{equation}
where $K=m \dot{x}^2/2$ is the kinetic energy and
$-V'(x)$ is the general pinning force of Eq.~\eqref{eq:genericpinningforce}.
Equation~\eqref{eq:phasespaceeqmotion}, which follows directly from Eq.~\eqref{eq:mainequation},
is only valid if $\dot{x}\geq 0$ but allow us to obtain the limit-cycle trajectories directly
in phase space $(x,\dot{x})$.
We set $\gamma=1$ and for a given $f<f_c^0$ we prepare a limit cycle of the bistable regime,
using suitable values for $m$ and initial conditions.
The smallest value of the mass, $m^*$, that makes the limit-cycle trajectory $\dot{x}(x)$ touch the
$\dot{x}=0$ axis in one point (and in all its periodic images) corresponds to the homoclinic orbit.
The magenta dashed line in Fig.~\ref{fig:phasespaceorbits}(b) illustrates one such homoclinic orbit.
The pair $(f,m)\equiv (f_c(m),m)$ found hence belongs to the homoclinic bifurcation line.
The critical mass $m_c$ can be determined from the vanishing of $f_c^0-f_c(m)$ as $m \to m_c$.
To solve Eq.~\eqref{eq:phasespaceeqmotion} numerically, for several values of $\Upsilon$ around the
standard value $\Upsilon=2$, we used the Runge-Kutta Fehlberg 78 method~\cite{BoostLibrary}.

\begin{figure}[tb]
\centering
\includegraphics[width=.9\columnwidth]{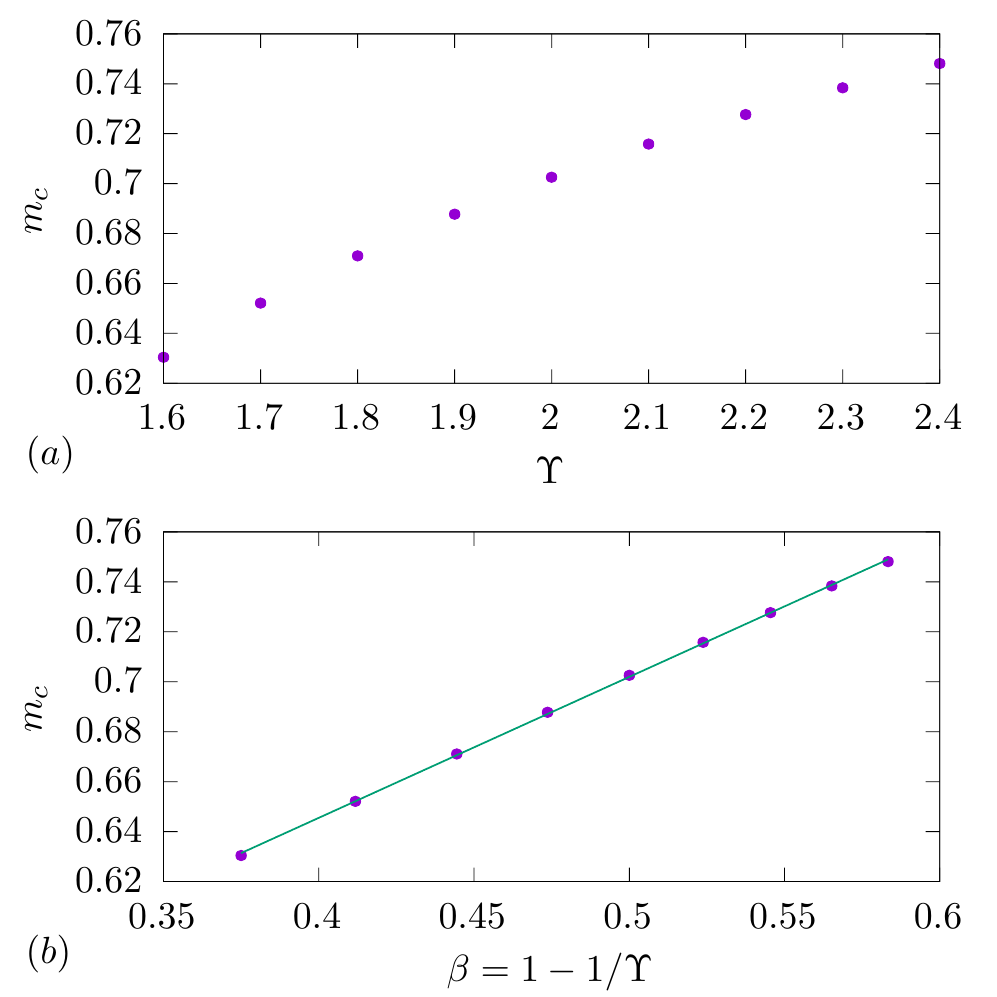}
\caption{
Critical mass $m_c$ vs. the normal form exponent $\Upsilon$ (a) and $\beta = 1-1/\Upsilon$ (b).
A fair linear fit is obtained in (b) in the full range of $\Upsilon$ shown in (a).
}
\label{fig:McVsUpsilon}
\end{figure}

\begin{figure}[tb]
\centering
\includegraphics[width=.9\columnwidth]{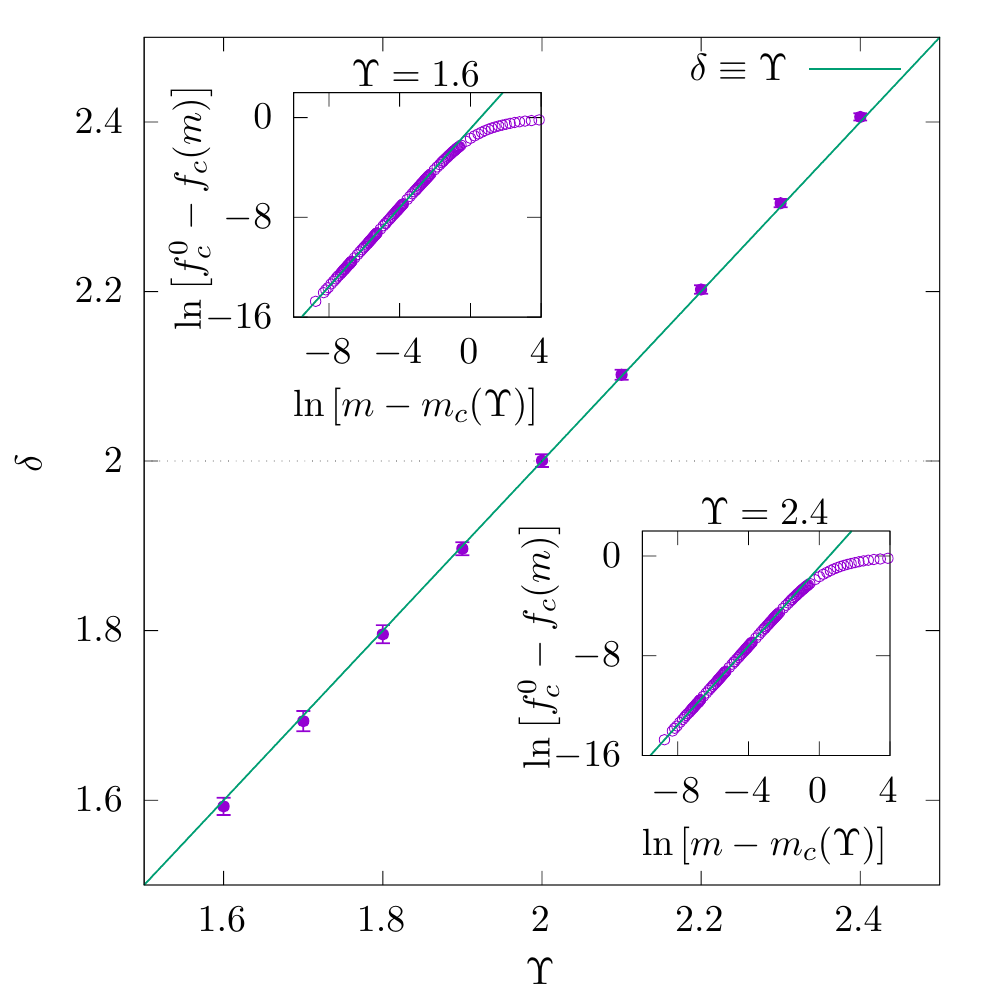}
\caption{
Critical exponent $\delta$, controlling the bistable driving force range $f_c^0-f_c(m)$,
showing that $\delta \approx \Upsilon$ (main figure).
Insets show two typical power-law fits $f_c^0-f_c(m) \approx \left[m-m_c(\Upsilon)\right]^{\delta}$
used to extract $\delta$ vs. $\Upsilon$.
}
\label{fig:deltaVsUpsilon}
\end{figure}

\begin{figure}[tb]
\centering
\includegraphics[width=.9\columnwidth]{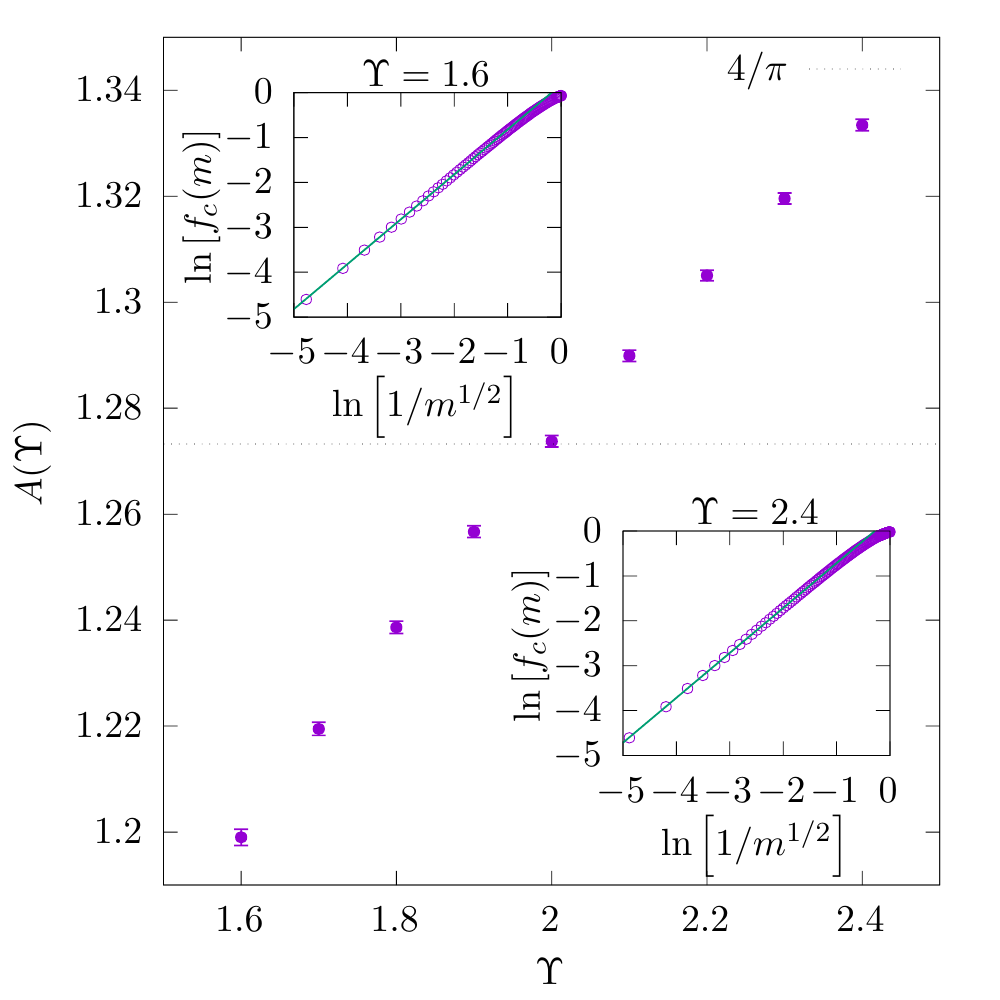}
\caption{
Large mass behavior of the critical force $f_c(m)$ vs. $\Upsilon$.
Insets: Fits to $f_c(m) \approx A(\Upsilon)/\sqrt{m}$.
Main figure: $A(\Upsilon)$ for a range of $\Upsilon$ around the standard $\Upsilon=2$
[or $-\cos(u)$ potential].
}
\label{fig:MelkinovVsUpsilon}
\end{figure}

In the insets of Fig.~\ref{fig:deltaVsUpsilon}, we show that
$f_c^0-f_c(m) \sim (m-m_c)^{\delta}$ as $m-m_c\to 0+$ for different values of $\Upsilon$,
with both $m_c$ and $\delta$ functions of $\Upsilon$,
as shown in Fig.~\ref{fig:McVsUpsilon} and Fig.~\ref{fig:deltaVsUpsilon} respectively.
As we can appreciate in Fig.~\ref{fig:deltaVsUpsilon}, $\delta$ is not an independent exponent, and
$\delta \approx \Upsilon$, as predicted analytically.
It is worth noting here that also depinning exponent $\beta$ is not an independent exponent,
since $\beta=1-1/\Upsilon$.
In other words, the bistable range is controlled by the normal form exponent
corresponding to the infinite-period bifurcation for $m>m_c$.
Summarizing,
\begin{align}
f_c^0-f_c(m) &\sim (m-m_c)^{\Upsilon}, \;\;\;\;\;\, m \lesssim m_c \\
v &\sim (f-f_c^0)^{1-1/\Upsilon}, \; m>m_c,\; f>f_c^0.
\end{align}
In Fig.~\ref{fig:McVsUpsilon} we show that $m_c$ is a nontrivial function of $\Upsilon$, for which
we do not have analytical prediction.
Interestingly, $m_c$ displays the behavior $m_c \approx a + b (1-1/\Upsilon)$ or
$m_c(\Upsilon) \approx m_c(2) + b (1/2-1/\Upsilon)$
with $b$ a positive constant in the neighborhood of the standard $\Upsilon=2$ case.
Finally, in Fig.~\ref{fig:MelkinovVsUpsilon}, we show that in the homoclinic bifurcation line
satisfies the following scaling form (see insets)
\begin{align}
f_c(m) \sim A(\Upsilon)/\sqrt{m}, \; \alpha \to 0
\end{align}
with $A(\Upsilon)$ a nontrivial prefactor (see main figure).
One recovers as expected the Guckenheimer and Holmes prediction $A(\Upsilon=2)= 4/\pi$, whose
derivation is detailed in Sec.~\ref{sec:gener-scal-theory}.
The low-damping scaling $f_c(m) \sim 1/\sqrt{m}$ (or $f_c^\alpha \sim \alpha$) is hence robust under
changes of $\Upsilon$, at variance with the  $m \to m_c$ critical behavior which displays the
$\Upsilon$-dependent exponent $\delta = \Upsilon$.
All these results agree with the analytical arguments made for the more general pinning force in
Sec.~\ref{sec:gener-scal-theory}.

\section{Conclusions}
\label{sec:conclusions}
We have studied the dynamical phase diagram as a function of the drive and damping of a massive
particle in a periodic potential.
The phase diagram consists of three different regimes, pinned, sliding and bistable, separated by
three bifurcation lines, each one identified with a different type of depinning transition on
driving.
We have obtained analytical descriptions of the homoclinic bifurcation line which separates the
bistable and the pinned regimes, both in the triple point and in the low damping limits.

The asymptotic behavior of the homoclinic bifurcation line presents interesting universal
features.
On one hand, for $\alpha-\alpha_c \to 0^+$, we find
$f_c^0-f_c(\alpha) \sim (\alpha_c - \alpha)^\Upsilon$,
with $\Upsilon$ representing an infinite family of periodic potentials solely characterized by the
normal form describing the shape of the periodic force near its minima.
The critical mass depends in a non critical way with $\Upsilon$.
For the cosine potential in particular, corresponding to $\Upsilon=2$, we were able to obtain
analytical estimates of the critical mass.
On the the other hand, for $\alpha \to 0$, we find $f_c(\alpha) \sim \mathcal{N}\alpha$ and obtain
an expression for $\mathcal{N}$.
This scaling result was already known for $\Upsilon=2$.
Nevertheless, we presented a physical argument which allows us to recover the
prediction by Guckenheimer and Holmes in this particular case~\cite{guckenheimer1983}.
Interestingly, this scaling result is more robust than the triple point scaling, as it appears to be
independent of the normal form exponent $\Upsilon$.
To the best of our knowledge many of these properties were not reported before, particularly
regarding the proximity of the triple point.

Using standard numerical methods, we have validated our analytical predictions.
In the case of the cosine potential, near the triple point,
the analytical result for $f_c^0-f_c(m)$ given by Eq.~\eqref{eq:scalingregimeclosetofc0exactcos}
is in excellent agreement with numerical data, down to the prefactor.
However, the numerical result for the critical mass is close but it does not match
the analytical prediction---which was an approximation---given that the critical mass is not a
universal property of the model.
For the generic normal forms parametrized by the exponent $\Upsilon$ we have also found a good
agreement on the $\Upsilon$-dependent predicted exponents.
In the latter case the critical mass follow a simple but nontrivial linear relation near
$\Upsilon=2$, $m_c \sim a + b \beta$,
with $\beta = 1 - 1/\Upsilon$ the velocity critical exponent for $m<m_c$.
Since we do not have an analytical prediction for this behavior, it would be interesting to tackle
this problem in the future.

The analytical estimates, which we reported and validated numerically, are obtained by
a nonstandard approach.
This consists in mapping exact static soliton solutions in a modified tilted washboard potential to
the homoclinic orbit,
which separates the bistable and the pinned regimes in the original dynamical model.
This approach appears to be a particularly useful alternative for an accurate description of
underdamped nonuniform oscillators driven near their triple point.

Regarding possible applications of our results to concrete physical systems, it would be interesting
to investigate 
how thermal fluctuations affect the dynamics, particularly near the triple point---either for the
model we considered, or for other nonlinear dynamics that present a coupling with
an inertia-like degree of freedom, e.g., in simple models of spintronic
devices~\cite{lecomte_depinning_2009,
stewart_e_barnes_and_jean-pierre_eckmann_and_thierry_giamarchi_and_vivien_lecomte_noise_2012}.
Vollmer and Risken used a functional continued fraction approach to study the small-damping limit in
presence of a noise
\cite{risken_brownian_1979,risken_low_1979,vollmer_bistability_1980,risken_fokker-planck_1996}
but it would be interesting to determine if it can describe the critical regime
close to the triple point.

\begin{acknowledgments}
We acknowledge the France-Argentina project ECOS-Sud No. A16E03. 
V.H.P. acknowledges hospitality at LIPhy and UGA, where this project was kick started.
V.L.~acknowledges support by the ERC Starting Grant No. 680275 MALIG, the ANR-18-CE30-0028-01 Grant
LABS and the ANR-15-CE40-0020-03 Grant LSD. A.B.K. acknowledges partial support from Grants No.
PICT2016-0069/FONCyT and No. 06/C578/UNCUYO from Argentina.
\end{acknowledgments}

\section{Appendix}
\label{sec:appendix}

\subsection{Implementation of the numerical integration}
From the computational point of view, this model does not present major difficulties when studying
the cosine potential.
Nevertheless, since the inertial term has a second order derivative, we employ a Verlet's
integration method to solve the dynamics.
We use a Leapfrog algorithm, performing the integration in two steps:
First, we calculate the velocity $v(t)$, as $v(t) \equiv \dot{x}$, and then the resulting
position, at each time step.
That means, at each time step $dt$ we update the velocity of the particle, as
\begin{equation}
\label{eq:leapfrog_next_v}
v(t+dt) = v(t) + \left[{F_{\cos}(x,t) - v(t)}\right]\, \frac{dt}{m},
\end{equation}
with $F_{\cos}(x,t)$ as defined on Eq.~\eqref{eq:defF0sin}.
Then we update the position by using the updated velocity,
\begin{equation}
\label{eq:leapfrog_next_x}
x(t+dt) = x(t)+ v(t+dt)\, dt.
\end{equation}

To guarantee the stability of this method, we use a zero-acceleration initial condition for the
particle, i.e., $\ddot{x} \equiv 0$.
For the initial position and velocity, we fix
\begin{equation}
\label{eq:simple_pend_initial_cond}
x(0) = 0, \qquad v(0) = F_{\cos}(0,0) = f.
\end{equation}
By using this initial condition, we study the dynamics of a particle under a range of driving
forces and different values for its mass.

\begin{figure}[tb]
\centering
\includegraphics[width=.9\columnwidth]{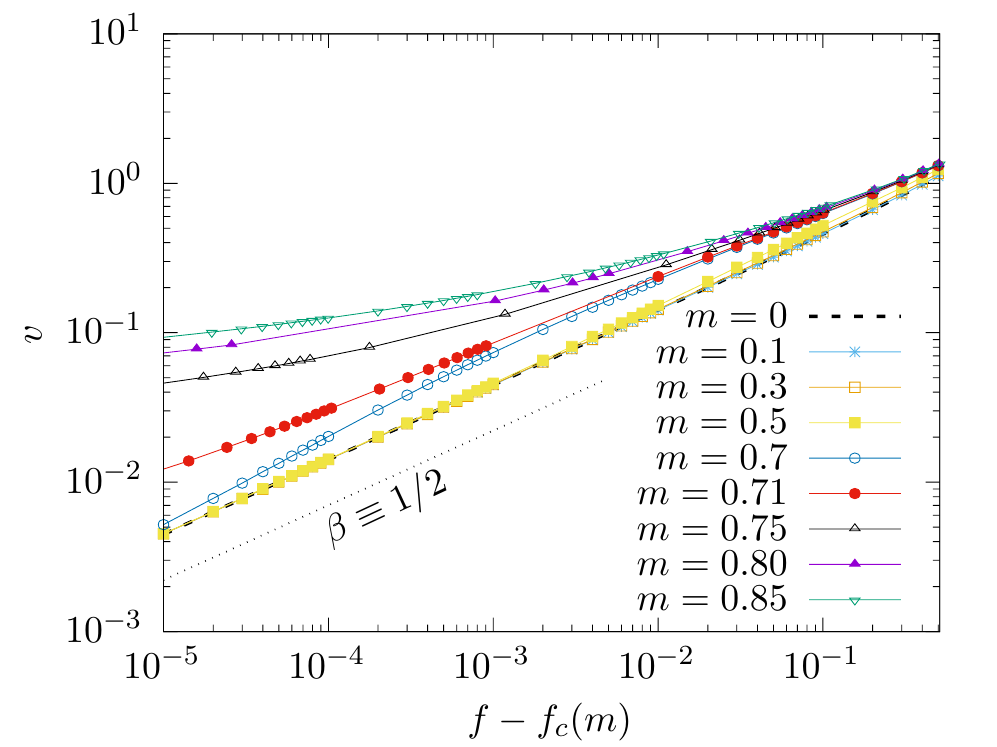}
\caption{
On the cosine potential, the steady-state time-averaged velocity, as a function of the force,
undergoes different depinning transitions depending on the mass of the particle.
For zero-mass, the black dashed line shows the analytical result, with $\beta=1/2$.
For masses above $m_c$, we can only see a mild tendency toward the regime $\beta\to0$
that describes the expected behavior $v\sim\left\lbrace\ln[f-f_c(m)]\right\rbrace^{-1}$.
}
\label{fig:v_vs_f_diffm}
\end{figure}

Close to the depinning point, we expect different behaviors, depending on the mass of the particle.
On the one hand, below the critical mass, the depinning exponent is $\beta = 1/2$.
On the other hand, above the critical mass, the steady-state time-averaged velocity undergoes an
abrupt transition, proportional to
$\left\lbrace\ln[f-f_c(m)]\right\rbrace^{-1}$~\cite[Sec.~8.5]{strogatz2018nonlinear}.
Hence, we expect $\beta\to0$ in this case.
However, the finite precision of a computer makes it very tedious to obtain such a
inverse-logarithmic behavior.
In Fig.~\ref{fig:v_vs_f_diffm}, we illustrate how the depinning transition close to the
critical mass is manifested in practice, on the velocity-force characteristics.
As the mass increases, above the critical one, we observe a tendency of the critical velocity
characteristics toward a regime where $\beta\to 0$ (i.e., toward a horizontal line)
but studying regimes with much smaller values of $f-f_c(m)$ would be required in order to observe
numerically the inverse-logarithmic behavior.

\subsection{Numerical method to find the critical mass}
\begin{figure}[tb]
\centering
\includegraphics[width=.9\columnwidth]{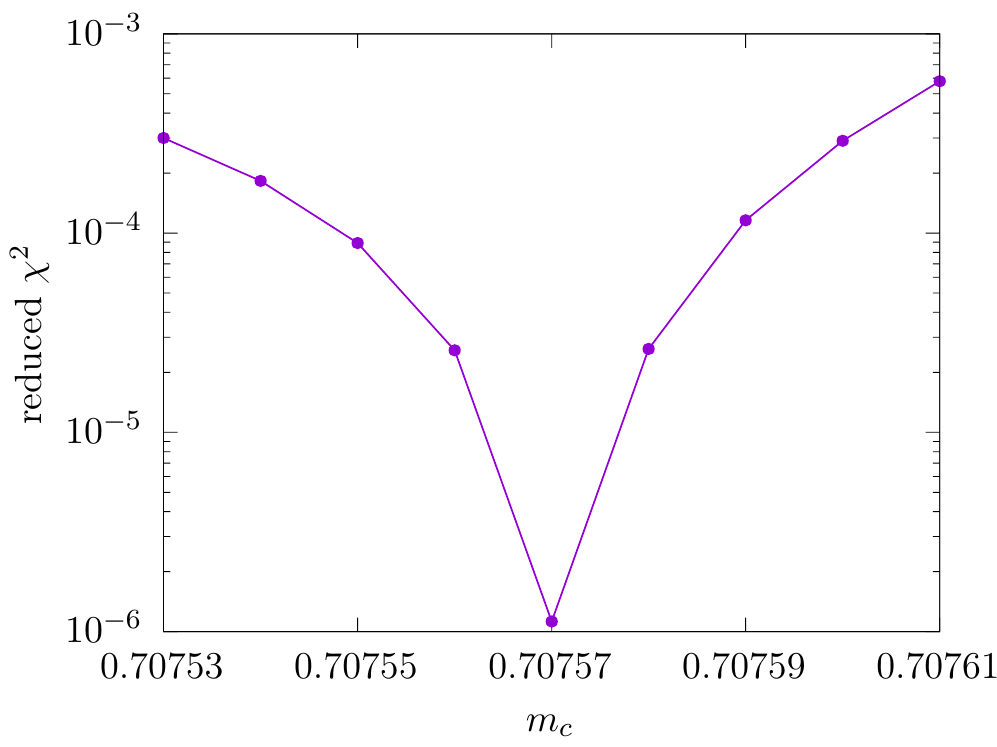}
\caption{
Using Eq.~\eqref{eq:power_law_m-mc},
we fit the exponent by proposing different values for the critical mass.
Here we show the resulting reduced chi-square for each value, as a test for the goodness of fit.
From the lowest value of the reduced reduced chi-square, we estimate $m_c=0.70757\pm0.00002$.
}
\label{fig:reduced_chisquared}
\end{figure}

From the steady-state time-averaged velocity as a function of the mass,
for a particle driven by $f=f_c^0$,
we can fit the critical mass and a critical exponent using Eq.~\eqref{eq:power_law_m-mc}.
Since one of the fitting parameters, $m_c$, is part of the argument of the power law,
the fitting method is not straightforward.
In our case, we fitted the data proposing different values of $m_c$,
using only the exponent as fitting parameter.
Besides the estimation for the latter,
for each $m_c$ we test the quality of the fit using the reduced chi-square.
Then, to estimate the critical mass, we simply choose the best fit,
as we show in Fig.~\ref{fig:reduced_chisquared}.

% \clearpage
\bibliography{references.bib}

\begin{thebibliography}{19}%
\makeatletter
\providecommand \@ifxundefined [1]{%
 \@ifx{#1\undefined}
}%
\providecommand \@ifnum [1]{%
 \ifnum #1\expandafter \@firstoftwo
 \else \expandafter \@secondoftwo
 \fi
}%
\providecommand \@ifx [1]{%
 \ifx #1\expandafter \@firstoftwo
 \else \expandafter \@secondoftwo
 \fi
}%
\providecommand \natexlab [1]{#1}%
\providecommand \enquote  [1]{``#1''}%
\providecommand \bibnamefont  [1]{#1}%
\providecommand \bibfnamefont [1]{#1}%
\providecommand \citenamefont [1]{#1}%
\providecommand \href@noop [0]{\@secondoftwo}%
\providecommand \href [0]{\begingroup \@sanitize@url \@href}%
\providecommand \@href[1]{\@@startlink{#1}\@@href}%
\providecommand \@@href[1]{\endgroup#1\@@endlink}%
\providecommand \@sanitize@url [0]{\catcode `\\12\catcode `\$12\catcode
  `\&12\catcode `\#12\catcode `\^12\catcode `\_12\catcode `\%12\relax}%
\providecommand \@@startlink[1]{}%
\providecommand \@@endlink[0]{}%
\providecommand \url  [0]{\begingroup\@sanitize@url \@url }%
\providecommand \@url [1]{\endgroup\@href {#1}{\urlprefix }}%
\providecommand \urlprefix  [0]{URL }%
\providecommand \Eprint [0]{\href }%
\providecommand \doibase [0]{http://dx.doi.org/}%
\providecommand \selectlanguage [0]{\@gobble}%
\providecommand \bibinfo  [0]{\@secondoftwo}%
\providecommand \bibfield  [0]{\@secondoftwo}%
\providecommand \translation [1]{[#1]}%
\providecommand \BibitemOpen [0]{}%
\providecommand \bibitemStop [0]{}%
\providecommand \bibitemNoStop [0]{.\EOS\space}%
\providecommand \EOS [0]{\spacefactor3000\relax}%
\providecommand \BibitemShut  [1]{\csname bibitem#1\endcsname}%
\let\auto@bib@innerbib\@empty
%</preamble>
\bibitem [{\citenamefont {Strogatz}(2018)}]{strogatz2018nonlinear}%
  \BibitemOpen
  \bibfield  {author} {\bibinfo {author} {\bibfnamefont {S.~H.}\ \bibnamefont
  {Strogatz}},\ }\href {\doibase 10.1201/9780429492563} {\emph {\bibinfo
  {title} {Nonlinear Dynamics and Chaos: With Applications to Physics, Biology,
  Chemistry, and Engineering}}},\ \bibinfo {edition} {2nd}\ ed.\ (\bibinfo
  {publisher} {CRC Press},\ \bibinfo {year} {2018})\BibitemShut {NoStop}%
\bibitem [{\citenamefont {Tinkham}(2004)}]{tinkham2004introduction}%
  \BibitemOpen
  \bibfield  {author} {\bibinfo {author} {\bibfnamefont {M.}~\bibnamefont
  {Tinkham}},\ }\href {https://store.doverpublications.com/0486435032.html}
  {\emph {\bibinfo {title} {Introduction to Superconductivity}}},\ \bibinfo
  {edition} {2nd}\ ed.\ (\bibinfo  {publisher} {Dover Publications},\ \bibinfo
  {year} {2004})\BibitemShut {NoStop}%
\bibitem [{\citenamefont {Barone}\ and\ \citenamefont
  {Patern{\`o}}(2005)}]{baronepaterno}%
  \BibitemOpen
  \bibfield  {author} {\bibinfo {author} {\bibfnamefont {A.}~\bibnamefont
  {Barone}}\ and\ \bibinfo {author} {\bibfnamefont {G.}~\bibnamefont
  {Patern{\`o}}},\ }\enquote {\bibinfo {title} {Voltage current
  characteristics},}\ in\ \href {\doibase 10.1002/352760278X.ch6} {\emph
  {\bibinfo {booktitle} {Physics and Applications of the Josephson Effect}}}\
  (\bibinfo  {publisher} {John Wiley \& Sons, Ltd},\ \bibinfo {year} {2005})\
  Chap.~\bibinfo {chapter} {6}, pp.\ \bibinfo {pages} {121--160},\ \bibinfo
  {edition} {1st}\ ed.\BibitemShut {Stop}%
\bibitem [{\citenamefont {Purrello}\ \emph {et~al.}(2017)\citenamefont
  {Purrello}, \citenamefont {Iguain}, \citenamefont {Kolton},\ and\
  \citenamefont {Jagla}}]{purrello2017}%
  \BibitemOpen
  \bibfield  {author} {\bibinfo {author} {\bibfnamefont {V.~H.}\ \bibnamefont
  {Purrello}}, \bibinfo {author} {\bibfnamefont {J.~L.}\ \bibnamefont
  {Iguain}}, \bibinfo {author} {\bibfnamefont {A.~B.}\ \bibnamefont {Kolton}},
  \ and\ \bibinfo {author} {\bibfnamefont {E.~A.}\ \bibnamefont {Jagla}},\
  }\href {\doibase 10.1103/PhysRevE.96.022112} {\bibfield  {journal} {\bibinfo
  {journal} {Phys. Rev. E}\ }\textbf {\bibinfo {volume} {96}},\ \bibinfo
  {pages} {022112} (\bibinfo {year} {2017})}\BibitemShut {NoStop}%
\bibitem [{\citenamefont {Stratonovich}(1965)}]{stratonovich_oscillator_1965}%
  \BibitemOpen
  \bibfield  {author} {\bibinfo {author} {\bibfnamefont {R.~L.}\ \bibnamefont
  {Stratonovich}},\ }in\ \href {\doibase 10.1016/B978-1-4832-3230-0.50026-2}
  {\emph {\bibinfo {booktitle} {Non-{Linear} {Transformations} of {Stochastic}
  {Processes}}}}\ (\bibinfo  {publisher} {Elsevier},\ \bibinfo {year} {1965})\
  pp.\ \bibinfo {pages} {269--282}\BibitemShut {NoStop}%
\bibitem [{\citenamefont {Ambegaokar}\ and\ \citenamefont
  {Halperin}(1969)}]{ambegaokar1969}%
  \BibitemOpen
  \bibfield  {author} {\bibinfo {author} {\bibfnamefont {V.}~\bibnamefont
  {Ambegaokar}}\ and\ \bibinfo {author} {\bibfnamefont {B.~I.}\ \bibnamefont
  {Halperin}},\ }\href {\doibase 10.1103/PhysRevLett.22.1364} {\bibfield
  {journal} {\bibinfo  {journal} {Phys. Rev. Lett.}\ }\textbf {\bibinfo
  {volume} {22}},\ \bibinfo {pages} {1364} (\bibinfo {year}
  {1969})}\BibitemShut {NoStop}%
\bibitem [{\citenamefont {Doussal}\ and\ \citenamefont
  {Vinokur}(1995)}]{ledoussal_1995}%
  \BibitemOpen
  \bibfield  {author} {\bibinfo {author} {\bibfnamefont {P.~L.}\ \bibnamefont
  {Doussal}}\ and\ \bibinfo {author} {\bibfnamefont {V.~M.}\ \bibnamefont
  {Vinokur}},\ }\href {\doibase 10.1016/0921-4534(95)00545-5} {\bibfield
  {journal} {\bibinfo  {journal} {Physica C: Superconductivity}\ }\textbf
  {\bibinfo {volume} {254}},\ \bibinfo {pages} {63} (\bibinfo {year}
  {1995})}\BibitemShut {NoStop}%
\bibitem [{\citenamefont {Scheidl}(1995)}]{scheidl_mobility_1995}%
  \BibitemOpen
  \bibfield  {author} {\bibinfo {author} {\bibfnamefont {S.}~\bibnamefont
  {Scheidl}},\ }\href {\doibase 10.1007/BF01307487} {\bibfield  {journal}
  {\bibinfo  {journal} {Z. Physik B - Condensed Matter}\ }\textbf {\bibinfo
  {volume} {97}},\ \bibinfo {pages} {345} (\bibinfo {year} {1995})}\BibitemShut
  {NoStop}%
\bibitem [{\citenamefont {Risken}(1996)}]{risken_fokker-planck_1996}%
  \BibitemOpen
  \bibfield  {author} {\bibinfo {author} {\bibfnamefont {H.}~\bibnamefont
  {Risken}},\ }\href {\doibase 10.1007/978-3-642-61544-3} {\emph {\bibinfo
  {title} {The {Fokker}-{Planck} Equation: Methods of Solution and
  Applications}}},\ \bibinfo {edition} {2nd}\ ed.,\ \bibinfo {series} {Springer
  Series in Synergetics}, Vol.~\bibinfo {volume} {18}\ (\bibinfo  {publisher}
  {Springer-Verlag Berlin Heidelberg},\ \bibinfo {year} {1996})\BibitemShut
  {NoStop}%
\bibitem [{\citenamefont {Bishop}\ and\ \citenamefont
  {Trullinger}(1978)}]{bishop1978}%
  \BibitemOpen
  \bibfield  {author} {\bibinfo {author} {\bibfnamefont {A.~R.}\ \bibnamefont
  {Bishop}}\ and\ \bibinfo {author} {\bibfnamefont {S.~E.}\ \bibnamefont
  {Trullinger}},\ }\href {\doibase 10.1103/PhysRevB.17.2175} {\bibfield
  {journal} {\bibinfo  {journal} {Phys. Rev. B}\ }\textbf {\bibinfo {volume}
  {17}},\ \bibinfo {pages} {2175} (\bibinfo {year} {1978})}\BibitemShut
  {NoStop}%
\bibitem [{\citenamefont {Levi}\ \emph {et~al.}(1978)\citenamefont {Levi},
  \citenamefont {Hoppensteadt},\ and\ \citenamefont
  {Miranker}}]{levi1978dynamics}%
  \BibitemOpen
  \bibfield  {author} {\bibinfo {author} {\bibfnamefont {M.}~\bibnamefont
  {Levi}}, \bibinfo {author} {\bibfnamefont {F.~C.}\ \bibnamefont
  {Hoppensteadt}}, \ and\ \bibinfo {author} {\bibfnamefont {W.}~\bibnamefont
  {Miranker}},\ }\href {\doibase 10.1090/qam/484023} {\bibfield  {journal}
  {\bibinfo  {journal} {Quarterly of Applied Mathematics}\ }\textbf {\bibinfo
  {volume} {36}},\ \bibinfo {pages} {167} (\bibinfo {year} {1978})}\BibitemShut
  {NoStop}%
\bibitem [{\citenamefont {Mel'nikov}(1963)}]{melnikov1963}%
  \BibitemOpen
  \bibfield  {author} {\bibinfo {author} {\bibfnamefont {V.~K.}\ \bibnamefont
  {Mel'nikov}},\ }\href {http://mi.mathnet.ru/eng/mmo/v12/p3} {\bibfield
  {journal} {\bibinfo  {journal} {Trudy Moskovskogo Matematicheskogo
  Obshchestva}\ }\textbf {\bibinfo {volume} {12}},\ \bibinfo {pages} {3}
  (\bibinfo {year} {1963})}\BibitemShut {NoStop}%
\bibitem [{\citenamefont {Guckenheimer}\ and\ \citenamefont
  {Holmes}(1983)}]{guckenheimer1983}%
  \BibitemOpen
  \bibfield  {author} {\bibinfo {author} {\bibfnamefont {J.}~\bibnamefont
  {Guckenheimer}}\ and\ \bibinfo {author} {\bibfnamefont {P.}~\bibnamefont
  {Holmes}},\ }\href {\doibase 10.1007/978-1-4612-1140-2} {\emph {\bibinfo
  {title} {Nonlinear oscillations, dynamical systems, and bifurcations of
  vector fields}}},\ \bibinfo {edition} {1st}\ ed.,\ \bibinfo {series} {Applied
  Mathematical Sciences}, Vol.~\bibinfo {volume} {42}\ (\bibinfo  {publisher}
  {Springer-Verlag New York},\ \bibinfo {year} {1983})\BibitemShut {NoStop}%
\bibitem [{\citenamefont {Anhert}\ and\ \citenamefont
  {Mulansky}(2015)}]{BoostLibrary}%
  \BibitemOpen
  \bibfield  {author} {\bibinfo {author} {\bibfnamefont {K.}~\bibnamefont
  {Anhert}}\ and\ \bibinfo {author} {\bibfnamefont {M.}~\bibnamefont
  {Mulansky}},\ }\href@noop {} {\enquote {\bibinfo {title} {Boost {C++}
  {L}ibraries/odeint},}\ }\bibinfo {howpublished}
  {\url{https://www.boost.org/doc/libs/1_73_0/libs/numeric/odeint}} (\bibinfo
  {year} {2015}),\ \bibinfo {note} {last revised: April 22, 2020}\BibitemShut
  {NoStop}%
\bibitem [{\citenamefont {Lecomte}\ \emph {et~al.}(2009)\citenamefont
  {Lecomte}, \citenamefont {Barnes}, \citenamefont {Eckmann},\ and\
  \citenamefont {Giamarchi}}]{lecomte_depinning_2009}%
  \BibitemOpen
  \bibfield  {author} {\bibinfo {author} {\bibfnamefont {V.}~\bibnamefont
  {Lecomte}}, \bibinfo {author} {\bibfnamefont {S.~E.}\ \bibnamefont {Barnes}},
  \bibinfo {author} {\bibfnamefont {J.-P.}\ \bibnamefont {Eckmann}}, \ and\
  \bibinfo {author} {\bibfnamefont {T.}~\bibnamefont {Giamarchi}},\ }\href
  {\doibase 10.1103/PhysRevB.80.054413} {\bibfield  {journal} {\bibinfo
  {journal} {Phys. Rev. B}\ }\textbf {\bibinfo {volume} {80}},\ \bibinfo
  {pages} {054413} (\bibinfo {year} {2009})}\BibitemShut {NoStop}%
\bibitem [{\citenamefont {Barnes}\ \emph {et~al.}(2012)\citenamefont {Barnes},
  \citenamefont {Eckmann}, \citenamefont {Giamarchi},\ and\ \citenamefont
  {Lecomte}}]{stewart_e_barnes_and_jean-pierre_eckmann_and_thierry_giamarchi_and_vivien_lecomte_noise_2012}%
  \BibitemOpen
  \bibfield  {author} {\bibinfo {author} {\bibfnamefont {S.~E.}\ \bibnamefont
  {Barnes}}, \bibinfo {author} {\bibfnamefont {J.-P.}\ \bibnamefont {Eckmann}},
  \bibinfo {author} {\bibfnamefont {T.}~\bibnamefont {Giamarchi}}, \ and\
  \bibinfo {author} {\bibfnamefont {V.}~\bibnamefont {Lecomte}},\ }\href
  {http://iopscience.iop.org/0951-7715/25/5/1427} {\bibfield  {journal}
  {\bibinfo  {journal} {Nonlinearity}\ }\textbf {\bibinfo {volume} {25}},\
  \bibinfo {pages} {1427} (\bibinfo {year} {2012})}\BibitemShut {NoStop}%
\bibitem [{\citenamefont {Risken}\ and\ \citenamefont
  {Vollmer}(1979{\natexlab{a}})}]{risken_brownian_1979}%
  \BibitemOpen
  \bibfield  {author} {\bibinfo {author} {\bibfnamefont {H.}~\bibnamefont
  {Risken}}\ and\ \bibinfo {author} {\bibfnamefont {H.~D.}\ \bibnamefont
  {Vollmer}},\ }\href {\doibase 10.1007/BF01323506} {\bibfield  {journal}
  {\bibinfo  {journal} {Z Physik B}\ }\textbf {\bibinfo {volume} {33}},\
  \bibinfo {pages} {297} (\bibinfo {year} {1979}{\natexlab{a}})}\BibitemShut
  {NoStop}%
\bibitem [{\citenamefont {Risken}\ and\ \citenamefont
  {Vollmer}(1979{\natexlab{b}})}]{risken_low_1979}%
  \BibitemOpen
  \bibfield  {author} {\bibinfo {author} {\bibfnamefont {H.}~\bibnamefont
  {Risken}}\ and\ \bibinfo {author} {\bibfnamefont {H.~D.}\ \bibnamefont
  {Vollmer}},\ }\href {\doibase 10.1016/0375-9601(79)90384-0} {\bibfield
  {journal} {\bibinfo  {journal} {Physics Letters A}\ }\textbf {\bibinfo
  {volume} {69}},\ \bibinfo {pages} {387} (\bibinfo {year}
  {1979}{\natexlab{b}})}\BibitemShut {NoStop}%
\bibitem [{\citenamefont {Vollmer}\ and\ \citenamefont
  {Risken}(1980)}]{vollmer_bistability_1980}%
  \BibitemOpen
  \bibfield  {author} {\bibinfo {author} {\bibfnamefont {H.~D.}\ \bibnamefont
  {Vollmer}}\ and\ \bibinfo {author} {\bibfnamefont {H.}~\bibnamefont
  {Risken}},\ }\href {\doibase 10.1007/BF01352745} {\bibfield  {journal}
  {\bibinfo  {journal} {Z Physik B}\ }\textbf {\bibinfo {volume} {37}},\
  \bibinfo {pages} {343} (\bibinfo {year} {1980})}\BibitemShut {NoStop}%
\end{thebibliography}%
\end{document}